\documentclass{jfm}

\usepackage{bm}
\usepackage{siunitx}
\usepackage{subcaption}
\usepackage{tikz}
\usetikzlibrary{
  arrows.meta,
  decorations.pathmorphing,
  decorations.markings,
  patterns,
  calc
}

\usepackage{xcolor}

\newcommand{\Relam}{Re_\lambda}
\newcommand{\Relamph}{Re_\lambda(\phi)}

\newcommand{\eps}{\varepsilon}
\newcommand{\epsph}{\varepsilon(\phi)}
\newcommand{\taueta}{\tau_\eta}

\newcommand{\Ceps}{C_\varepsilon}
\newcommand{\xh}{x/h}
\newcommand{\Uo}{U_0}
\newcommand{\Uti}{\tilde{U}}

\newcommand{\sigV}{\sigma_V}

\newcommand{\sigRPP}{\sigma_\mathrm{RPP}}
\newcommand{\St}{\mathit{St}}

\newcommand{\taup}{\tau_p}

\newcommand{\vph}{\phi}

\lefttitle{M.~Ahmad et al.}

\title{Preferential concentration in a periodically forced particle-laden turbulent planar jet}

\author{Mohammad Ahmad\aff{1}\aff{2}
        ,
        Pierre Bragança\aff{1}
        ,
        Christophe Cuvier\aff{1}
        \and
        Mart\'in Obligado\aff{1}\aff{3}}

\affiliation{
  \aff{1}Univ. Lille, CNRS, ONERA, Arts et Metiers Institute of Technology,
  Centrale Lille, UMR 9014-LMFL-Laboratoire de M\'ecanique des Fluides de
  Lille - Kamp\'e de F\'eriet, F-59000 Lille, France
  \aff{2}Icam School of Engineering, Lille campus, 6 rue Auber, B.P. 10079,
  59016, Lille, France
  \aff{3}Institut universitaire de France (IUF), Paris, France
}

\corresau{M.~Ahmad, \email{mohammad.ahmad@icam.fr}}

\begin{document}

\maketitle

\begin{abstract}
Periodic forcing is common in particle-laden jets, yet its influence on the
spatial evolution of turbulence and inertial-particle clustering remains
poorly understood. We investigate a turbulent planar jet at two Reynolds
numbers, approximately 4950 and 9900 based on the nozzle height, subjected
to low-frequency sinusoidal forcing. Hot-wire anemometry and high-speed
particle imaging are used to characterise the carrier turbulence and
particle organisation. The forcing frequency is approximately three orders
of magnitude below the preferred instability frequency of the jet.
Phase-resolved measurements characterise the carrier turbulence, while
Voronoi tessellations quantify preferential concentration of dilute
inertial droplets.

Despite its low frequency and moderate amplitude, the forcing substantially
modifies the turbulence while leaving the time-averaged mean-flow development
essentially unchanged. The integral scale is reduced by up to $40\%$,
while forced and unforced jets retain the same time-averaged relationship
between the dimensionless dissipation coefficient and the Taylor-scale
Reynolds number. Within the forcing cycle, the turbulent velocity
fluctuation and Taylor microscale vary approximately in opposite phase,
leaving the Taylor-scale Reynolds number within $\pm10\%$ of its cycle
mean despite appreciable variations in dissipation and the Kolmogorov
time scale.

The dispersed phase is even more sensitive to this unsteadiness. Forcing
increases the mean preferential concentration and produces phase-dependent
variations in Voronoi clustering intensity reaching $21$--$49\%$ of
the cycle mean. Clustering is strongest when the Kolmogorov time scale is
largest, corresponding to the smallest particle Stokes number.
Consequently, markedly different clustering states occur at similar
instantaneous Taylor-scale Reynolds numbers. These results show that weak
temporal modulation can strongly modify turbulence scales and particle
organisation while leaving the mean transporting flow almost unchanged,
suggesting a route towards controlling preferential concentration in
particle-laden turbulent flows.
\end{abstract}


\section{Introduction}
\label{sec:intro}

Turbulent jets carrying suspended droplets or particles arise in a wide
range of engineering and environmental flows, including fuel injection
\citep{ReitzDiwakar1987}, spray drying \citep{MujumdarHandbook2014},
aerosol drug delivery \citep{NewmanPRD2017}, and atmospheric dispersion
associated with respiratory emissions and other transient releases
\citep{Bourouiba2014,Wang2021,Chong2021}. In many of these applications, the flow is
not statistically stationary but is instead generated in successive pulses
or under periodically varying operating conditions. As a result, both the
carrier turbulence and the dispersed phase evolve continuously in response
to the imposed forcing.

The dynamics of these two phases are intrinsically coupled. Turbulence
controls the transport, mixing and preferential concentration of inertial
particles, while the particle distribution depends on the evolving state
of the carrier flow. Understanding the response of particle-laden jets
therefore requires first establishing how the carrier turbulence itself
evolves under periodic forcing.

Despite the prevalence of pulsed particle-laden jets, this coupled problem
remains poorly understood. Previous studies have generally examined either
the evolution of the carrier turbulence or the dynamics of inertial
particles separately, whereas relatively little is known about their
combined response in a spatially developing turbulent jet. In particular,
the instantaneous turbulent state produced by periodic forcing, and its
relationship with the phase-resolved response of inertial particles, remain
poorly characterised.

Single-phase turbulent planar jets are among the canonical free-shear flows
used to investigate the spatial development of turbulence. Owing to their
simple geometry and well-defined self-similar far field, they have served
for decades as a benchmark for studying entrainment, turbulent mixing and
the growth of turbulence scales in spatially developing shear flows
\citep{Townsend1976,Pope2000}. Beyond an initial development
region, the jet approaches a statistically self-preserving state in which
the centreline velocity decreases approximately as $(x/h)^{-1/2}$ (where $h$ is the nozzle height), the jet
half-width grows linearly with downstream distance, and suitably normalised
mean and turbulent quantities collapse onto nearly universal profiles
\citep{Gutmark1976}.

These classical similarity scalings are commonly associated with the
equilibrium assumption that the turbulent kinetic energy dissipation rate $\varepsilon$ scales as
$\varepsilon \sim u_\mathrm{rms}^3/L$, where $u_\mathrm{rms}$ is the
root-mean-square turbulent velocity fluctuation and $L$ is the integral
length scale. This is equivalent to assuming that the normalised
dissipation coefficient, $C_\varepsilon=\frac{\varepsilon L}{u_\mathrm{rms}^3}$,
remains approximately constant throughout the self-preserving region
\citep{Townsend1976,Pope2000}. The resulting self-preserving state has
established the planar jet as a reference configuration for examining
turbulence evolution through entrainment, spreading and streamwise
development.

While the self-similar behaviour of the mean flow is now well understood,
the downstream evolution of the turbulence itself remains an active area of
research. As ambient fluid is progressively entrained into the jet, the
energy-containing eddies grow, the integral and Taylor scales evolve, and
the dissipation adjusts to the changing local flow conditions. Together,
these quantities provide complementary descriptions of the turbulence and
of the transfer of kinetic energy across scales, making them natural
diagnostics for characterising the downstream development of the jet
\citep{Deo2008,Xie_Zhang_Xiong_Zhou_2024}.

Recent studies have suggested that the equilibrium assumption underlying
the classical similarity solution may not hold exactly in all spatially
developing turbulent flows, including planar jets \citep{Cafiero2019,Vassilicos2015}.
Related departures from equilibrium dissipation scaling have been reported
in wakes and decaying grid turbulence
\citep{Valente2012,Dairay2015}. Regardless of their
interpretation, these observations demonstrate that the evolution of the
turbulence scales and dissipation provides a sensitive description of the
downstream development of the flow.

Introducing periodic forcing at the nozzle superimposes a temporal
modulation onto this inherently spatial evolution. In contrast with the
extensive literature devoted to statistically stationary planar jets,
comparatively little is known about how weak, low-frequency forcing modifies
the downstream development of the turbulence. Previous investigations of
forced jets and free shear layers have focused primarily on the coherent
structures generated by the Kelvin--Helmholtz instability, including
preferred-mode excitation \citep{CrowChampagne1971}, vortex pairing
\citep{ZamanHussain1980}, and entrainment enhancement
\citep{HoHuerre1984}. By comparison, the response of the turbulence scales
throughout the spatial development of the jet, and particularly their
instantaneous evolution through a forcing cycle, has received considerably
less attention. Characterising this unsteady evolution provides the
fluid-mechanical framework required to interpret the response of an
embedded dispersed phase.

The relevance of this unsteadiness becomes particularly apparent when considering inertial particles. Particles suspended in turbulent flows do not in general
distribute uniformly \citep{ToschiBodenschatz2009}. Instead, they preferentially accumulate in specific
regions of the flow, forming clusters whose characteristic size and
intensity depend on both particle inertia and the structure of the carrier
turbulence \citep{Maxey1987,Bec2007}.

The dynamics of sub-Kolmogorov inertial particles, such as those considered
in the present work, are commonly characterised by the Stokes number,
$\St=\tau_p/\tau_\eta$, defined as the ratio between the particle response
time $\tau_p$ and the Kolmogorov time scale $\tau_\eta$ of the carrier
turbulence. Preferential concentration is strongest for
$\St=O(1)$, where the finite response of the particles to rapidly
fluctuating velocity gradients gives rise to intense clustering
\citep{BraggIrelandCollins2015,BrandtColetti2022}. Although several
physical mechanisms contribute to this behaviour, they establish a close
connection between inertial clustering and the local structure and time
scales of the carrier turbulence.

Among the various diagnostics available to quantify preferential
concentration, Voronoi tessellation has become one of the most widely used
because it provides a simple, parameter-free characterisation of the local
particle concentration \citep{Monchaux2010}. Its robustness has led to
widespread application in both experimental and numerical studies of
particle-laden turbulence \citep{Monchaux2012,falkinhoff2020preferential}.
Voronoi-based approaches have been applied across a broad range of canonical
turbulent flows, including grid turbulence \citep{Saw2008,Obligado2014},
homogeneous isotropic turbulence
\citep{Monchaux2017,baker2017coherent}, and swirling flows
\citep{Angriman2022b,Ferran2023}. In the present work, Voronoi tessellation
is employed to quantify the instantaneous and phase-resolved evolution of
preferential concentration, providing access to both the clustering
intensity and the characteristic cluster size throughout the forcing cycle.

Periodic forcing introduces an additional characteristic time scale into
the problem through the forcing period $T_f=1/f$, or equivalently the
forcing Strouhal number $St_h=fh/U_0$, where $U_0$ is the bulk exit velocity. Forcing at $St_h=O(10^{-1})$ is well known
to excite the preferred Kelvin--Helmholtz mode of turbulent jets
\citep{HoHuerre1984}, producing organised vortex dynamics such as vortex
pairing \citep{CrowChampagne1971} and modifications of entrainment
\citep{ZamanHussain1980}. By contrast, the present experiments are
conducted at $St_h\sim10^{-4}$, approximately three orders of magnitude
below the preferred-mode regime. The forcing therefore acts as a slow
modulation of an already developed turbulent jet rather than as an
excitation of the primary shear-layer instability. This regime provides a
framework for isolating how slow temporal forcing modifies the downstream
evolution and instantaneous properties of developed turbulence and,
consequently, the dynamics of the dispersed phase.

Because the particle response is governed by
$\St=\tau_p/\tau_\eta$, any temporal evolution of the Kolmogorov time scale,
$\tau_\eta=(\nu/\varepsilon)^{1/2}$, produces a corresponding variation of
the effective particle inertia. Even particles with fixed physical
properties therefore experience a phase-dependent Stokes number,
$\St(\phi)=\tau_p/\tau_\eta(\phi)$, whose evolution is dictated by the
instantaneous state of the carrier turbulence. Whether preferential
concentration follows these instantaneous variations, exhibits a delayed
response, or develops more complex phase-dependent dynamics remains an open
question.

Direct numerical simulations have suggested that temporally varying
turbulence can modify inertial clustering relative to statistically
stationary conditions \citep{Zapata2024}. Experimental evidence, however,
remains scarce. To the best of our knowledge, no previous experimental
study has systematically characterised phase-resolved preferential
concentration in a periodically forced turbulent planar jet. More
generally, the relationship between the instantaneous properties of an
unsteady carrier turbulence and the corresponding state of inertial-particle
clustering remains poorly established.

The present study addresses these questions through complementary
single-phase and particle-laden measurements in a periodically forced
turbulent planar air jet. The first objective is to determine how weak
periodic forcing modifies the spatial development of the carrier
turbulence, considering not only its time-averaged properties but also the
phase-resolved evolution of its turbulence scales and dissipation. This
provides a direct characterisation of the unsteadiness and of the
instantaneous turbulent states sampled throughout the forcing cycle. The
second objective is to determine how the inertial-particle field responds
to these states. Low-frequency sinusoidal forcing
($Re_h\approx4950$ and $9900$, $f=0.125$--$0.25$\,Hz, corresponding to
values of $St_h$ up to $7.5\times10^{-4}$) is applied at the nozzle, and
the carrier turbulence is characterised over the full streamwise
development using phase-resolved hot-wire anemometry with triple
decomposition \citep{Hussain1970}. High-speed particle imaging combined
with Voronoi tessellation is then employed at the same streamwise location
to quantify the phase evolution of preferential concentration.

The measurements show that weak periodic forcing preserves the
time-averaged self-similar development of the planar jet while
systematically modifying the evolution of its turbulence scales. Although
the forced and unforced jets share the same time-averaged mean-flow
structure and the same time-averaged relationship between the normalised
dissipation coefficient and the Taylor-scale Reynolds number, forcing
modifies the downstream evolution of the turbulence scales and introduces
a reproducible phase dependence in the carrier turbulence. Periodic forcing
therefore reorganises the evolution of the turbulence without altering the
canonical time-averaged development of the jet.

The particle measurements further show that preferential concentration
responds directly to these temporal variations of the carrier turbulence.
Phase-resolved clustering follows the evolution of the small-scale
turbulence through the forcing cycle rather than the nearly invariant
Taylor-scale Reynolds number, demonstrating that the instantaneous
clustering state cannot be parameterised by $Re_\lambda$ alone under
periodically varying conditions. Taken together, these measurements
demonstrate how periodic forcing modifies both the spatial and instantaneous
evolution of turbulence in a canonical planar jet, and how this unsteadiness
governs the instantaneous clustering of inertial particles.
\section{Experimental set-up}
\label{sec:setup}

\subsection{Jet facility and sinusoidal forcing}

The experiments were performed in a planar air jet built to the same design
as that of \citet{Cafiero2019}, whose main features are summarised here for
completeness.  A centrifugal blower feeds ambient air into a large settling
chamber that damps the residual unsteadiness introduced by the motor; the
flow is then conditioned by two successive screen and straightener stages and
accelerated through a smooth contraction of area ratio close to $8$.  This
contraction brings the centreline turbulence intensity down to below $1\%$ at
entry to the nozzle, which is a letterbox slit of height
$h = 15\,\mathrm{mm}$ and span $H = 465\,\mathrm{mm}$, giving an aspect ratio
$H/h = 31$.  To secure a clean top-hat exit profile and avoid tripping the
boundary layers at the lips, the two long edges of the slit are rounded with
a fillet radius $r = 2h$, as recommended by \citet{Deo2008}.  Beyond the exit
the jet discharges into the quiescent laboratory air inside a rectangular
perspex enclosure measuring $100h \times 58h \times 31h$
($1500 \times 870 \times 465\,\mathrm{mm}^3$), which preserves the
two-dimensionality of the mean flow and restricts spanwise entrainment.

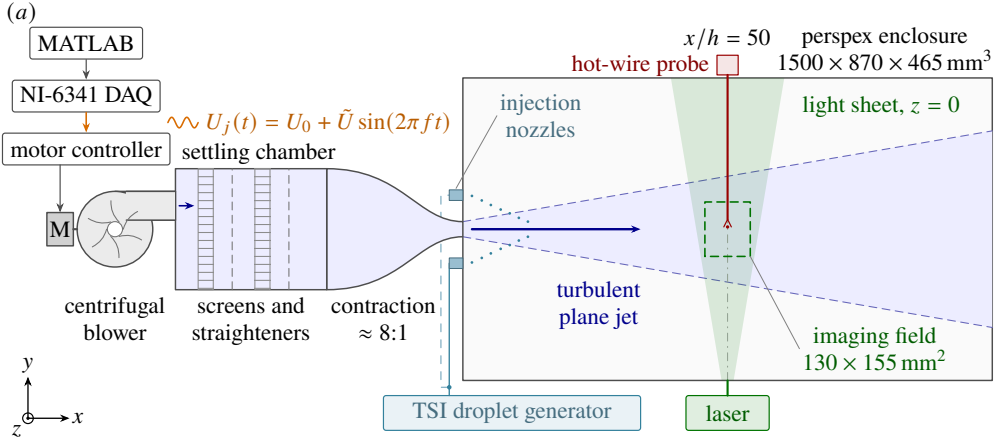
\begin{figure}
  \centering
  \begin{tikzpicture}[
    >={Stealth[length=3.6pt,width=2.8pt]},
    line cap=round, line join=round,
    font=\footnotesize,
    wall/.style={draw=black!70, line width=0.55pt},
    thinwall/.style={draw=black!45, line width=0.4pt},
    box/.style={draw=black!70, line width=0.45pt, fill=white, rounded corners=1pt,
                minimum height=0.42cm, inner xsep=3pt, inner ysep=1.5pt, align=center},
    lbl/.style={inner sep=1pt, align=center},
    lead/.style={draw=black!50, line width=0.3pt},
    dim/.style={draw=black!55, line width=0.3pt, {Stealth[length=2.6pt,width=2pt]}-{Stealth[length=2.6pt,width=2pt]}},
  ]

\colorlet{airfill}{blue!7}
\colorlet{jetline}{blue!55!black}
\colorlet{lasercol}{green!45!black}
\colorlet{hwacol}{red!60!black}
\colorlet{dropcol}{cyan!55!black}
\colorlet{sigcol}{orange!85!black}

\node[anchor=north west, inner sep=0pt] at (-6.05,3.0) {$(a)$};

\filldraw[wall, fill=black!2] (0,-2.0) rectangle (7.0,2.0);
\fill[airfill] (0,0.1) -- (7.0,1.3) -- (7.0,-1.3) -- (0,-0.1) -- cycle;
\draw[jetline!60, densely dashed, line width=0.4pt] (0,0.1) -- (7.0,1.3);
\draw[jetline!60, densely dashed, line width=0.4pt] (0,-0.1) -- (7.0,-1.3);
\draw[->, jetline, line width=0.9pt] (0.12,0) -- (2.35,0);

\draw[black!35, dash dot, line width=0.35pt] (3.5,-2.0) -- (3.5,2.0);
\fill[lasercol, opacity=0.13] (3.42,-2.0) -- (2.75,2.0) -- (4.25,2.0) -- (3.58,-2.0) -- cycle;
\draw[lasercol, line width=0.6pt, densely dashed] (3.2,-0.36) rectangle (3.8,0.36);
\filldraw[draw=lasercol, fill=lasercol!12, line width=0.45pt, rounded corners=1pt] (2.95,-2.66) rectangle (4.05,-2.2);
\node[lasercol!80!black] at (3.5,-2.43) {laser};
\draw[lasercol, line width=0.8pt] (3.5,-2.2) -- (3.5,-2.0);
\filldraw[draw=hwacol, fill=hwacol!10, line width=0.45pt] (3.36,2.04) rectangle (3.64,2.3);
\draw[hwacol, line width=0.8pt] (3.5,2.04) -- (3.5,0.13);
\draw[hwacol, line width=0.45pt] (3.5,0.13) -- (3.45,0.03) (3.5,0.13) -- (3.55,0.03);
\fill[hwacol] (3.5,0.03) circle (0.022);

\filldraw[draw=dropcol, fill=dropcol!8, line width=0.45pt, rounded corners=1pt] (-1.05,-2.66) rectangle (2.35,-2.2);
\node[dropcol!70!black] at (0.65,-2.43) {TSI droplet generator};
\draw[dropcol, line width=0.5pt] (-0.18,-2.2) -- (-0.18,-0.45);
\draw[dropcol!70, line width=0.4pt, dashed] (-0.18,-2.09) -- (-0.29,-2.09) -- (-0.29,0.45) -- (-0.18,0.45);
\fill[dropcol] (-0.18,-2.09) circle (0.03);
\foreach \s in {1,-1}{
  \filldraw[draw=dropcol!70!black, fill=dropcol!45, line width=0.35pt]
      (-0.18,0.38*\s) rectangle (0,0.52*\s);
  \foreach \px/\py in {0.12/0.44, 0.25/0.38, 0.37/0.32, 0.50/0.26, 0.62/0.21, 0.74/0.15, 0.86/0.10}
      {\fill[dropcol] (\px,\py*\s) circle (0.018);}
}

\filldraw[wall, fill=airfill] (-1.8,0.8) .. controls (-0.8,0.8) and (-0.6,0.1) .. (0,0.1)
    -- (0,-0.1) .. controls (-0.6,-0.1) and (-0.8,-0.8) .. (-1.8,-0.8) -- cycle;
\filldraw[wall, fill=airfill] (-3.8,-0.8) rectangle (-1.8,0.8);
\foreach \xa in {-3.5,-2.75}{
  \fill[pattern=horizontal lines, pattern color=black!45] (\xa,-0.8) rectangle (\xa+0.2,0.8);
  \draw[thinwall] (\xa,-0.8) rectangle (\xa+0.2,0.8);
  \draw[black!55, densely dashed, line width=0.4pt] (\xa+0.45,-0.8) -- (\xa+0.45,0.8);
}
\draw[->, jetline, line width=0.6pt] (-3.75,0.31) -- (-3.55,0.31);
\filldraw[wall, fill=black!8] (-4.6,0.12) rectangle (-3.8,0.5);
\filldraw[wall, fill=black!8] (-4.6,0) circle (0.5);
\fill[black!8] (-4.45,0.13) rectangle (-3.81,0.49);
\draw[black!70, line width=0.55pt] (-4.45,0.5) -- (-3.8,0.5) (-4.45,0.12) -- (-3.8,0.12);
\foreach \a in {0,60,...,300}
  {\draw[black!55, line width=0.4pt] ($(-4.6,0)+(\a:0.11)$) to[bend left=22] ($(-4.6,0)+(\a+38:0.38)$);}
\filldraw[black!55, fill=white, line width=0.4pt] (-4.6,0) circle (0.1);
\filldraw[wall, fill=black!18] (-5.5,-0.22) rectangle (-5.15,0.22);
\node at (-5.325,0) {M};
\draw[black!70, line width=0.8pt] (-5.15,0) -- (-5.1,0);

\node[box] (mat) at (-4.98,2.46) {MATLAB};
\node[box] (daq) at (-4.98,1.76) {NI-6341 DAQ};
\node[box] (ctl) at (-4.98,1.06) {motor controller};
\draw[->, black!70, line width=0.45pt] (mat.south) -- (daq.north);
\draw[->, sigcol, line width=0.6pt] (daq.south) -- (ctl.north);
\draw[->, black!70, line width=0.45pt] (-5.325,0.85) -- (-5.325,0.22);
\draw[sigcol, line width=0.6pt, domain=0:1, samples=40, variable=\t]
    plot ({-3.9+0.4*\t},{1.41+0.07*sin(720*\t)});
\node[anchor=west, sigcol!85!black, inner sep=1pt] at (-3.44,1.41) {$U_j(t)=\Uo+\Uti\sin(2\pi f t)$};

\node[lbl] at (-2.7,1.0) {settling chamber};
\node[lbl] at (-4.6,-0.86) [anchor=north] {centrifugal\\blower};
\node[lbl] at (-2.8,-0.86) [anchor=north] {screens and\\straighteners};
\node[lbl] at (-1.05,-0.86) [anchor=north] {contraction\\$\approx 8{:}1$};

\node[lbl, anchor=south east] at (7.05,2.06) {perspex enclosure\\$1500\times870\times465\,\mathrm{mm}^3$};
\node[lbl, anchor=south] at (3.5,2.33) {$x/h=50$};
\node[lbl, anchor=east, hwacol!85!black] at (3.3,2.17) {hot-wire probe};
\node[lbl, anchor=west, lasercol!80!black] at (4.45,1.62) {light sheet, $z=0$};
\node[lbl, anchor=west, lasercol!80!black] (fov) at (4.45,-1.58) {imaging field\\$130\times155\,\mathrm{mm}^2$};
\draw[lead] (fov.west) -- (3.8,-0.36);
\node[lbl, anchor=west, jetline] at (1.2,-1.0) {turbulent\\plane jet};
\node[lbl, anchor=west, dropcol!70!black] (inj) at (0.45,1.5) {injection\\nozzles};
\draw[lead] (inj.west) -- (-0.09,0.54);

\begin{scope}[shift={(-5.75,-2.5)}]
  \draw[->, line width=0.5pt] (0,0) -- (0.55,0) node[right, inner sep=1pt] {$x$};
  \draw[->, line width=0.5pt] (0,0) -- (0,0.55) node[above, inner sep=1pt] {$y$};
  \draw[line width=0.45pt] (0,0) circle (0.07); \fill (0,0) circle (0.022);
  \node[anchor=north east, inner sep=1pt] at (-0.04,-0.04) {$z$};
\end{scope}

\end{tikzpicture}
  \caption{Planar jet facility and measurement configuration (not to scale).
  Top view in the plane $z=0$. The blower speed follows
  (\ref{eq:forcing}) through the waveform sent by MATLAB and the NI-6341
  card to the motor controller. Droplets from the TSI generator enter the
  enclosure through two injection nozzles, one on each side of the jet; the
  dashed feed line runs below the enclosure. The hot-wire probe and the
  imaging field are both centred on the jet axis at $x/h=50$; the two
  campaigns were conducted independently.
  The origin $x=y=z=0$ is located at the centre of the nozzle exit plane.}
  \label{fig:setup}
\end{figure}

A Cartesian coordinate system is defined with its origin at the geometric
centre of the nozzle exit plane: $x$ is the streamwise direction, $y$ the
cross-stream direction along the narrow dimension $h$, and $z$ the spanwise
direction along $H$.  All measurements reported here were performed in the
central plane $z = 0$.  A schematic of the facility and measurement
configuration is shown in figure~\ref{fig:setup}.

Sinusoidal unsteadiness was imposed by modulating the blower speed so that
the jet exit velocity followed
\begin{equation}
  U_j(t) = \Uo + \Uti\sin(2\pi f t),
  \label{eq:forcing}
\end{equation}
where $\Uo$ is the mean exit velocity and $\Uti$ and $f$ are the modulation
amplitude and frequency.  The sinusoidal waveform was generated digitally in
MATLAB, converted to an analogue voltage signal by a National Instruments
NI-6341 data acquisition card, and supplied directly to the motor
controller.  The control system was operated open-loop; preliminary tests
using a Pitot tube placed in the plenum chamber confirmed that the jet exit
velocity followed the commanded waveform accurately for the frequencies and
amplitudes investigated.  The modulation amplitude is limited by the
rotational inertia of the blower motor, which constrains the achievable
amplitude at higher frequencies; the resulting amplitudes $\Uti/\Uo \leq 20\%$
place all conditions in a moderate, small-amplitude regime.
This same inertial roll-off also sets the upper bound of the accessible
forcing range: above $f \approx 0.25$\,Hz the commanded sinusoid could no
longer be faithfully reproduced at the nozzle exit, which fixes the
highest frequency investigated here.
%


\subsection{Single-phase flow: Hot-wire anemometry campaign}

Single-point velocity measurements were performed using hot-wire anemometry (HWA) used on constant-temperature
anemometry (CTA) mode. The system consists of a Dantec Dynamics Streamline 90N10
frame equipped with a 90C10 single-channel CTA module and a single-wire probe.
The sensing element is a platinum-plated tungsten wire of diameter
$5\,\mu\mathrm{m}$ and sensing length $1.25\,\mathrm{mm}$
(length-to-diameter ratio 250), operated at an overheat ratio of 1.8.  The
probe was mounted in a custom support with a NACA~0015 airfoil profile to
minimise flow disturbance from the stem, and positioned using a manual linear
traversing system with a resolution of $0.02\,\mathrm{mm}$ in $y$ and
$1\,\mathrm{mm}$ in $x$ and $z$.

The analogue voltage signal was conditioned (offset $1.625\,\mathrm{V}$,
gain $\times 16$) and digitised at $F_s = 50\,\mathrm{kHz}$ with 16-bit
resolution using the NI-6341 card. Prior to digitisation the
conditioned signal was analogue low-pass filtered below the Nyquist
frequency $F_s/2 = 25\,\mathrm{kHz}$ to suppress aliasing. The wire sensing
length $\ell_w = 1.25\,\mathrm{mm}$ sets the small-scale spatial resolution;
the corresponding ratio $\ell_w/\eta$ at $x/h = 50$ is quantified in
\S\,\ref{sec:methods:hwa}.  The sampling rate $F_s = 50\,\mathrm{kHz}$ lies
well above the Kolmogorov frequency $f_\eta = U_c/(2\pi\eta)$ at $x/h = 50$
for every condition, so the acquisition temporally resolves the smallest
turbulent time scales without ambiguity.  Because both the local mean
velocity $U_c$ and the dissipation rate are largest near the nozzle, $f_\eta$
attains its highest values at the most upstream stations; even there it
remains below the Nyquist frequency $F_s/2 = 25\,\mathrm{kHz}$, so the
temporal resolution is adequate over the full measurement domain and not only
at $x/h = 50$.  The complementary spatial resolution $\ell_w/\eta$, quantified
at $x/h = 50$ in \S\,\ref{sec:methods:hwa}, is coarsest at these upstream
stations for the same reason.  Record lengths were
$T_\mathrm{acq} = 60\,\mathrm{s}$ for the steady baseline cases and
$120\,\mathrm{s}$ for the forced cases, yielding at least 1000 integral-scale
turnover times and no fewer than 15 complete forcing cycles per record.  The
probe was calibrated in situ against a Pitot-static tube connected to a
high-precision manometer (resolution $0.01\,\mathrm{Pa}$) at the start and
end of each session, fitting approximately 10 velocity points, including
the zero-velocity limit, to King's Law.

Axial measurements were acquired at 11 streamwise stations from $x/h = 0$ to
$90$.  Cross-stream profiles were measured at stations
$x/h = 5, 10, 30, 50, 70$ and $90$, from $y/h \approx -4$ to at least
$1.5\,\delta(x)$, where $\delta(x)$ is the local jet half-width defined in
\S\,\ref{sec:methods}.  Spanwise uniformity was confirmed by comparing profiles at $z/h = 0$
and $z/h = \pm 10$ at a representative downstream station.

\subsection{Two-phase flow: High-speed particle imaging campaign}
\label{sec:setup:imaging}

Water droplets produced by a high-pressure droplet generator were introduced into the
jet through two injection nozzles located upstream of the measurement
section, one on each side of the jet (figure~\ref{fig:setup}).  All particle-imaging measurements were
performed at the fixed streamwise station $x/h = 50$, co-located with the HWA
centreline measurement point.  Images were acquired with a LaVision Imager MX
sCMOS camera at a frame rate of $10\,\mathrm{Hz}$, with a spatial resolution
of $60.7\,\mu\mathrm{m}\,\mathrm{px}^{-1}$ and a field of view of
approximately $130 \times 155\,\mathrm{mm}$ centred on the jet axis.  For each
experimental condition, $10\,000$ frames were recorded, providing
approximately 250 and 125 complete forcing cycles for the $f = 0.25\,\mathrm{Hz}$
and $f = 0.125\,\mathrm{Hz}$ conditions, respectively.

Each frame yields the instantaneous two-dimensional positions of the droplets
in the measurement plane; no velocity or size information is extracted from
these images.  The sole output of the imaging campaign is the Voronoi cell
area distribution and its standard deviation $\sigV$, computed as described
in \S\,\ref{sec:methods}.
Since the imaging is not used to size individual droplets, the droplet size
distribution was measured separately by shadowgraphy at the outlet of one
injection nozzle. The Sauter mean diameter is $D_{32}=30.9\,\mu\mathrm{m}$,
which gives a particle response time
$\tau_p = \rho_p D_{32}^{2}/(18\rho_f\nu) \approx 2.91\,\mathrm{ms}$ and the
particle Stokes numbers $\St=\tau_p/\taueta$ listed in
table~\ref{tab:conditions}.
The droplet liquid volume
fraction delivered by the generator
is $\phi_v \approx 1.0\times 10^{-5}$
at $\Uo = 5\,\mathrm{m\,s^{-1}}$ and $\approx 5.1\times 10^{-6}$ at
$\Uo = 10\,\mathrm{m\,s^{-1}}$.  The corresponding mass loading
$\phi_m \lesssim 10^{-2}$ is conventionally regarded as dilute, and we
adopt the one-way coupling assumption throughout.  In the regime map
of \citet{Elghobashi1994}, a volume fraction $\phi_v \sim 10^{-5}$ lies near
the boundary between the one-way and two-way coupling regimes; moreover, that
map was established for homogeneous turbulence, so the extent to which
one-way coupling holds locally in a free-shear flow, where the droplets
cluster strongly and the effective local loading can substantially exceed
its global value, remains an open question.

The two campaigns were conducted in separate, independent sessions; the
particle-imaging campaign was performed in the same facility under identical
forcing conditions as the HWA campaign, with the measurement volume
co-located at $x/h = 50$.

\subsection{Experimental conditions}

The six experimental conditions are summarised in table~\ref{tab:conditions}.
Two mean exit velocities $\Uo = 5$ and $10\,\mathrm{m\,s^{-1}}$ were studied,
corresponding to jet Reynolds numbers $Re_h = \Uo h/\nu \approx 4950$ and
$9900$.  For each velocity, a steady unforced baseline and two forced
conditions were investigated.  Case labels follow the convention: the
prefix S (steady, unforced) or T (periodically forced,
with the digit giving the forcing period in seconds) identifies the flow
state, and the numeral after the dash gives $\Uo$ in $\mathrm{m\,s^{-1}}$;
for example, T4-5 denotes $T = 4\,\mathrm{s}$ ($f = 0.25\,\mathrm{Hz}$)
at $\Uo = 5\,\mathrm{m\,s^{-1}}$.  The resulting forcing Strouhal numbers
$St_h = fh/\Uo$ lie in the range $1.88\text{--}7.50\times10^{-4}$, confirming
that the forcing period far exceeds the jet advection time $h/\Uo$; the
time-averaged mean flow is consequently quasi-steady, while the imposed
forcing nonetheless produces discernible unsteady effects, as verified in
\S\,\ref{sec:singlephase}.

\begin{table}
  \centering
  \caption{Summary of experimental conditions.
  $Re_h \equiv \Uo h/\nu$; $St_h \equiv fh/\Uo$.
  $\langle\Relam\rangle$ is the time-averaged Taylor-scale Reynolds number
  at $x/h = 50$ on the centreline, obtained from HWA.
  $\St \equiv \taup/\langle\taueta\rangle$ is the particle Stokes number
  based on the Sauter diameter ($\taup \approx 2.91\,\mathrm{ms}$,
  \S\,\ref{sec:setup:imaging}) and on the cycle-averaged Kolmogorov time scale
  measured at $x/h=50$; it varies through the forcing cycle
  (\S\,\ref{sec:twophase:phase}). It is distinct from the forcing
  Strouhal number $St_h$.
  Both campaigns (HWA along the full jet axis and high-speed particle
  imaging at $x/h = 50$) were conducted for all six conditions.}
  \label{tab:conditions}
  \begin{tabular}{lrrrrrccc}
    \toprule
    Case &
    $\Uo$ &
    $Re_h$ &
    $f$ &
    $T$ &
    $|\tilde{U}|$ &
    $St_h$ &
    $\St$ &
    $\langle\Relam\rangle$ \\
    &
    (m\,s$^{-1}$) &
    &
    (Hz) &
    (s) &
    (m\,s$^{-1}$) &
    &
     &
    at $x/h=50$ \\
    \midrule
    S-5   & 5  & 4\,950 & 0     & $\infty$ & 0   & 0                   & 0.56 & $\approx 120$ \\
    T4-5  & 5  & 4\,950 & 0.250 & 4        & 0.6 & $7.50\times10^{-4}$ & 0.55 & $\approx 97$  \\
    T8-5  & 5  & 4\,950 & 0.125 & 8        & 1.0 & $3.75\times10^{-4}$ & 0.59 & $\approx 93$  \\[3pt]
    S-10  & 10 & 9\,900 & 0     & $\infty$ & 0   & 0                   & 2.02 & $\approx 212$ \\
    T4-10 & 10 & 9\,900 & 0.250 & 4        & 0.6 & $3.75\times10^{-4}$ & 2.02 & $\approx 179$ \\
    T8-10 & 10 & 9\,900 & 0.125 & 8        & 1.0 & $1.88\times10^{-4}$ & 2.05 & $\approx 178$ \\
    \bottomrule
  \end{tabular}
\end{table}

\section{Data analysis}
\label{sec:methods}

\subsection{Hot-wire signal pre-processing}
\label{sec:methods:hwa}

The calibrated CTA signal is converted from the temporal to the spatial
domain using Taylor's frozen-turbulence hypothesis,
\begin{equation}
  k_1 = \frac{2\pi\,f_1}{U_c},
  \label{eq:taylors}
\end{equation}
where $f_1$ is the temporal frequency and $U_c$ is the local mean velocity
at the measurement station. At $x/h=50$, the turbulence intensity
$u_\mathrm{rms}/U_c$ does not exceed $12\%$ for any condition studied,
which lies within the commonly accepted range for the application of
Taylor's hypothesis in free-shear flows \citep{Pope2000}.

The sensing wire has a physical length
$l=1.25\,\mathrm{mm}$. At $x/h=50$, the Kolmogorov scale is
$\eta\approx0.3$--$0.5\,\mathrm{mm}$, yielding
$l/\eta\approx4$, at the upper end of the well-resolved range for
single-wire measurements. Single-wire hot-wire anemometry has previously
been shown to provide reliable estimates of dissipation and unsteady
cascade dynamics through the Kármán--Howarth framework
\citep{Obligado2019,Zheng2023}, motivating its use here for the
phase-resolved analysis.

\subsection{Triple decomposition and phase-resolved statistics}
\label{sec:methods:td}

The periodically forced velocity field contains contributions from the
time-averaged jet, the coherent response imposed by the forcing, and the
broadband turbulent fluctuations. To separate these three components we
adopt the triple decomposition introduced by \citet{Hussain1970},
\begin{equation}
U(t)=\overline{U}+\widetilde{U}(t)+u(t),
\label{eq:triple}
\end{equation}
where $\overline{U}$ denotes the time-averaged velocity,
$\widetilde{U}(t)$ the coherent periodic motion phase-locked to the
forcing, and $u(t)$ the remaining stochastic turbulent fluctuations.
Unlike the classical Reynolds decomposition, this formulation explicitly
isolates the deterministic forcing response from the turbulent component,
allowing the latter to be analysed independently throughout the forcing
cycle.

The forcing frequency is identified from the Welch power spectral density
of the raw velocity signal as the dominant spectral peak, which exceeds
the broadband turbulent background by more than
$30\,\mathrm{dB}$ for every forced case. The coherent component
$\widetilde{U}(t)$ is then extracted using a zero-phase narrow-band
Butterworth filter centred on the forcing frequency. A relative bandwidth
of $5\,\%$ (corresponding to a $-3\,\mathrm{dB}$ cutoff at $0.95f$)
provides sufficient separation between the coherent spectral line and the
surrounding turbulent continuum while avoiding phase distortion of the
periodic signal. The turbulent fluctuation is subsequently obtained as the
residual,
\[
u(t)=U(t)-\overline{U}-\widetilde{U}(t).
\]

\noindent The instantaneous forcing phase is defined as
\begin{equation}
\phi(t)=2\pi ft+\phi_0,
\label{eq:phase}
\end{equation}
where the constant phase offset $\phi_0$ is selected so that
$\phi\in[0,\pi)$ corresponds to the decelerating half-cycle and
$\phi\in[\pi,2\pi)$ to the accelerating half-cycle. Phase-resolved
statistics are obtained by partitioning each forcing period into
$N_\phi=18$ equally spaced bins of width
$\Delta\phi=20^\circ$. This choice provides sufficient phase resolution
while retaining adequate statistical convergence within each bin, and is
consistent with the phase- and ensemble-averaging strategies adopted in
previous studies of organised and periodically forced turbulent flows
\citep{Hussain1970, Zheng2023, Ahmad2026}.

Each acquisition lasts $120\,\mathrm{s}$, corresponding to
30 complete forcing periods for the T4 cases and 15 periods for the T8
cases. Consequently, each phase bin contains 30 and 15 statistically
independent realisations, respectively. The unforced reference cases are
processed identically but without phase binning.

Within each phase bin, one-dimensional energy spectra
$E_{11}(k_1;\phi)$ are estimated from the turbulent component
$u(t)$ using Welch's method
($N_{\mathrm{FFT}}=16384$, Hanning window, $25\,\%$ overlap). Computing
the spectra from the phase-conditioned turbulent signal enables all
derived turbulence quantities, including the dissipation rate, Taylor
microscale and integral scale, to be evaluated as functions of forcing
phase. Equivalent spectra obtained from the complete turbulent record are
used for comparisons with the statistically stationary quantities.

\subsection{Turbulence scales and dissipation coefficient}
\label{sec:methods:scales}

The phase-resolved dissipation rate is obtained from the isotropic relation
\begin{equation}
  \epsph = 15\nu \int_0^{\infty}
           k_1^2\,E_{11}(k_1;\phi)\,\mathrm{d}k_1.
  \label{eq:eps}
\end{equation}
The spectrum is integrated up to the wavenumber at which the signal reaches
the noise floor; beyond this point an exponential tail $A\exp(-k_1/B)$,
fitted to the last resolved decade of wavenumbers, is used to extrapolate the
integrand.  The Taylor microscale $\lambda(\phi)$ and Reynolds number
$\Relamph$ follow from
\begin{equation}
  \lambda(\phi) = \sqrt{\frac{15\nu\,u_\mathrm{rms}^2(\phi)}{\epsph}},
  \qquad
  \Relamph = \frac{u_\mathrm{rms}(\phi)\,\lambda(\phi)}{\nu}.
  \label{eq:lambda_relam}
\end{equation}
The integral length scale $L$ is computed from the temporal autocorrelation of the
turbulent signal:
\begin{equation}
  L(\phi) = U_c \int_0^{\tau_0} R_{uu}(\tau;\phi)\,\mathrm{d}\tau,
  \label{eq:L}
\end{equation}
where $\tau_0$ is the lag at which
$R_{uu}$ first falls to $1/e$, and the integral of the autocorrelation
(an integral time scale) is converted to a length via Taylor's
frozen-turbulence hypothesis, $L = U_c\,\mathcal{T}$ with $\mathcal{T} =
\int_0^{\tau_0} R_{uu}\,\mathrm{d}\tau$; the conversion uses the same local mean
$U_c$ as equation~\eqref{eq:taylors}, so that $L$ carries units of length.  This estimate is
cross-checked against the first zero-crossing of $R_{uu}$ following
\citet{Mora2020}. While we do not report results for the latter, all trends are conserved between both definitions of $L$.
The normalised dissipation coefficient is
\begin{equation}
  \Ceps(\phi) = \frac{\epsph\,L(\phi)}{u_\mathrm{rms}^3(\phi)}.
  \label{eq:ceps}
\end{equation}
In the standard Kolmogorov-like cascade $\Ceps$ is approximately constant; the
high-Reynolds non-equilibrium regime is identified by $\Ceps \sim \Relam^{-1}$
\citep{Vassilicos2015}.
In practice, the empirical exponent $b$ in $\Ceps \propto \Relam^{b}$
can sit between these two limits, with measured values $b \in [-1,\,0]$
reported across canonical flows \citep{Goto2016}; the exponent measured
here is reported in \S\,\ref{sec:singlephase:noneq}.

\subsection{Droplet detection}
\label{sec:methods:detection}

Each raw camera frame is processed in three stages to extract the
instantaneous two-dimensional positions of the droplets in the measurement
plane. First, the pixel-wise mean over the entire 10\,000-frame sequence is
computed and subtracted from each frame, removing static reflections, sensor
non-uniformity and residual illumination gradients. The background-subtracted
frame is then binarised with a fixed intensity threshold, and the resulting
connected components are filtered by their equivalent circular diameter.
This diameter is that of the recorded droplet image, not the
physical droplet: at the present magnification a droplet of diameter
$D_{32}\approx31\,\mu\mathrm{m}$ (about half a pixel geometrically) is
broadened by diffraction and
out-of-focus blur to a blob several pixels across, so a genuine droplet
occupies well above two pixels. The lower cut therefore removes single- and
sub-pixel intensity spikes (residual sensor and shot noise) rather than the
droplets themselves. Connected components smaller than two pixels
($\approx 120\,\mu\mathrm{m}$ in image space) are rejected
as sub-resolution noise, and those larger than approximately 40 pixels
($\approx 2.4\,\mathrm{mm}$) as non-droplet artefacts. The intensity-weighted
centroid of each surviving connected component finally defines one droplet
position, and the sole output for frame~$k$ is the set of two-dimensional
positions $\{(x_i,y_i)\}_{i=1}^{N_k}$: no velocity or size information
is retained.

\subsection{Voronoi tessellation and clustering statistic}
\label{sec:voronoi}

Preferential concentration is quantified using the Voronoi tessellation of
the instantaneous droplet positions \citep{Monchaux2010}.  For each frame $k$,
the detected positions are tessellated into Voronoi cells; any cell whose
boundary intersects the image frame is discarded to avoid area inflation at
the edges.  The area $\mathcal{A}_{i,k}$ of each surviving interior cell is
normalised by the global mean area,
\begin{equation}
  \overline{\mathcal{A}}
    = \frac{1}{\displaystyle\sum_k N_k}
      \sum_k \sum_{i=1}^{N_k} \mathcal{A}_{i,k},
  \label{eq:Amean}
\end{equation}
where $N_k$ is the number of interior cells in frame $k$ and the outer sum
runs over all 10\,000 frames.  Normalising by the global mean (rather than a
per-image mean) preserves the absolute number-density information: a locally
dense cluster manifests as cells with $\mathcal{A}_{i,k}/\overline{\mathcal{A}} \ll
1$ regardless of the total particle count in that frame.

The instantaneous clustering statistic is the standard deviation of the
normalised cell areas,
\begin{equation}
  \sigV(t_k)
    = \mathrm{std}\!\left(\frac{\mathcal{A}_{i,k}}{\overline{\mathcal{A}}}\right).
  \label{eq:sigV}
\end{equation}

Voronoi tessellation has become a standard tool for quantifying
preferential concentration in particle-laden turbulence because it provides
a parameter-free description of the local particle concentration
\citep{Monchaux2010,Monchaux2012,Monchaux2017}. For a Random Poisson Process (RPP), the probability density function of
the normalised Voronoi-cell area is well approximated by the universal
Gamma distribution derived by \citet{FerencNeda2007}. In two
dimensions, this reference distribution is characterised by a standard
deviation $\sigma_{\mathrm{RPP}}=0.53$, independently of the number of
particles \citep{Monchaux2010}. Consequently, values
$\sigma_V>\sigma_{\mathrm{RPP}}$ indicate departures from complete spatial
randomness and provide a global measure of preferential concentration. In consequence, and for a given set of boundary conditions, the larger the departure from the RPP value, the stronger the degree of
preferential concentration.

Besides this global metric, Voronoi tessellation also provides an objective
criterion for identifying individual clusters. Following
\citet{Monchaux2010}, particles whose Voronoi-cell areas are smaller than
the intersection between the measured and RPP Voronoi-area distributions
are classified as belonging to clusters. Although the present work focuses
on the global statistic $\sigma_V$, this criterion has become the standard
reference for Voronoi-based analyses of inertial-particle clustering.

The forcing phase $\phi(t_k)$, extracted from the triple decomposition
described in \S\,\ref{sec:methods:td}, is assigned to each imaging frame, from which phase-resolved
clustering statistics $\sigV(\phi)$ are obtained. Phase-averaged values
are obtained by sorting and averaging the
phase series $\{\sigV(\phi)\}$ into the same $N_\phi = 18$ bins of
$\Delta\phi = 20^\circ$ as the HWA analysis.

\section{Results: single-phase flow}
\label{sec:singlephase}

\subsection{Mean-flow self-similarity and invariance under forcing}
\label{sec:singlephase:meanflow}

\begin{figure}
  \includegraphics[width=\textwidth]{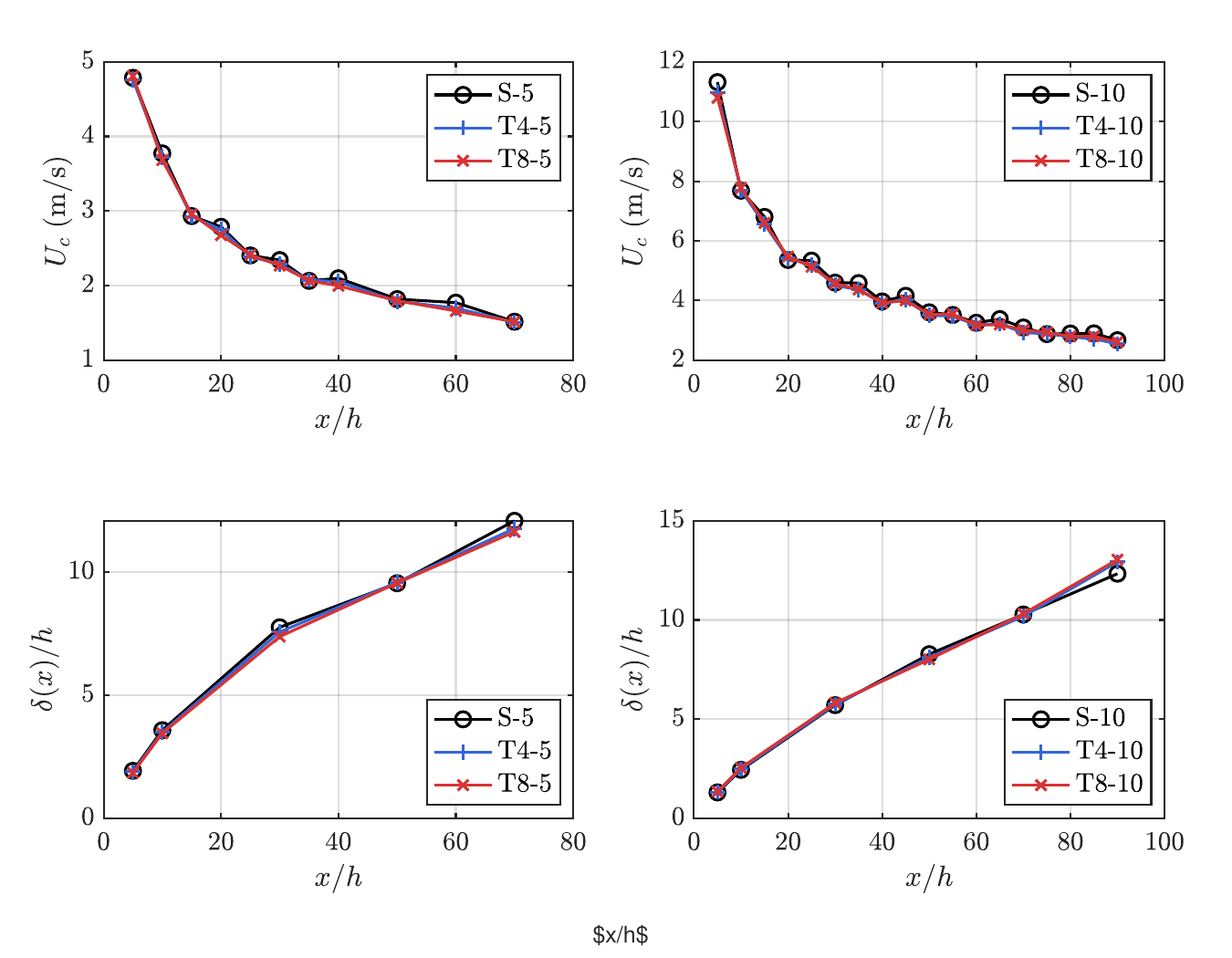}
  \caption{Time-averaged mean-flow development for all six experimental
  conditions. $(a,b)$~Centreline velocity $U_c(x/h)$ and
  $(c,d)$~normalised half-width $\delta(x/h)/h$ for exit velocities
  $\Uo = 5\,\mathrm{m\,s^{-1}}$ $(a,c)$ and
  $10\,\mathrm{m\,s^{-1}}$ $(b,d)$. All forced conditions collapse onto
  the unforced baseline at every station.}
  \label{fig:meanflow}
\end{figure}

The time-averaged centreline velocity $U_c(x)$ and jet half-width $\delta(x)$,
defined here as the transverse distance from the centreline to the point where
the mean velocity falls to half its centreline value ($U=U_c/2$), are presented
for all six experimental conditions in figure~\ref{fig:meanflow}. For both exit
velocities, $U_c$ exhibits a decay compatible with $(x/h)^{-1/2}$, while
$\delta$ grows approximately linearly for $x/h \gtrsim 10$. These trends are
consistent with the expected far-field self-similar development of a planar
jet \citep{Gutmark1976,Pope2000,Deo2008}, in accordance with the classical
self-preserving theory of free-shear flows
\citep{Goldstein1938,Townsend1976}.

The mean-flow scalings alone, however, do not provide a sufficiently sensitive
criterion to distinguish between different dissipation regimes. In the
self-similar framework of \citet{Cafiero2019}, a dissipation scaling
characterised by $m\in(0,1]$ leads to
\begin{equation}
    U_c \propto (x-x_0)^{-(m+1)/(4m+2)},
    \qquad
    \delta \propto (x-x_0)^{(m+1)/(2m+1)},
    \label{eq:cafiero_mean_scaling}
\end{equation}
where $x_0$ is the virtual origin. Over this range of $m$, the predicted
centreline decay exponent varies only between $1/3$ and $1/2$. Moreover, the
presence of the virtual origin introduces an additional degree of freedom
when fitting the spatial evolution over a finite streamwise range, making it
difficult to discriminate reliably between these relatively close scaling
exponents from the mean-flow data alone. The dissipation coefficient therefore
provides a more appropriate diagnostic of the underlying scaling regime. As
shown in \S\,\ref{sec:singlephase:noneq}, the measured dependence
$\Ceps \propto \Relam^{b}$, with $b=-m$, yields an empirical value
$m\approx0.58$. The corresponding mean-flow scaling remains compatible,
within the accessible streamwise range and experimental scatter, with the
apparent $U_c\propto x^{-1/2}$ behaviour observed in
figure~\ref{fig:meanflow}.

Interestingly, the three conditions at each exit velocity, comprising the
steady and both forced cases, are indistinguishable from one another across
all 11 measurement stations from $x/h=5$ to $90$. This collapse is not
limited to the centreline: the cross-stream velocity profiles (not shown in this work)
and the half-width evolution likewise confirm that the time-averaged
self-similar structure of the mean flow is preserved under forcing.

This mean-flow invariance is expected on dimensional grounds: the forcing
Strouhal number $St_h \sim 10^{-4}$ lies roughly three orders of magnitude
below the preferred mode of the jet shear-layer instability
\citep{CrowChampagne1971}, so the periodic modulation is too slow to interact
with the large-scale vortex dynamics responsible for mean-flow distortion.
From an experimental standpoint, this invariance is significant because it
ensures that the differences in turbulence statistics reported below cannot
be attributed to modifications of the time-averaged mean-flow development.
The six conditions therefore evolve within a common mean-flow background,
allowing the effect of periodic forcing on the turbulence itself to be
isolated. We now examine this effect through the downstream evolution of the
turbulence intensity and characteristic length scales.

\subsection{Single-phase turbulence in the steady baseline}
\label{sec:singlephase:steady}

\begin{figure}
  \includegraphics[width=\textwidth]{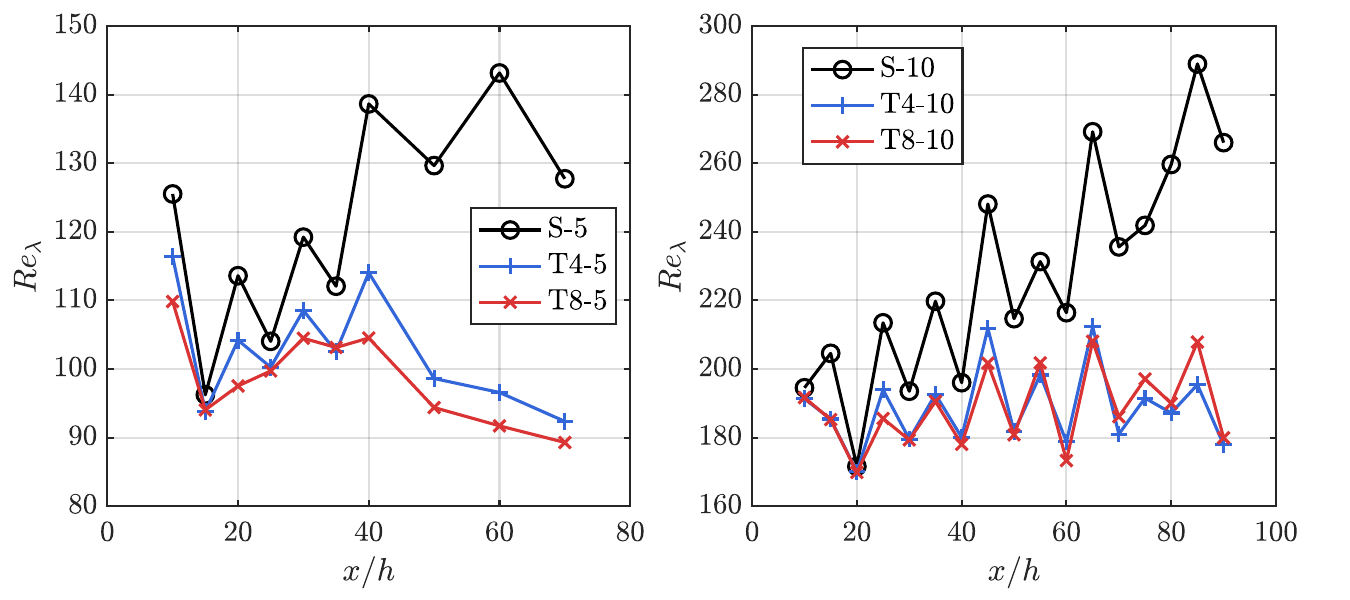}
  \caption{Taylor-scale Reynolds number $\Relam$ along the jet centreline
  for all six experimental conditions.
  $(a)$~$\Uo = 5\,\mathrm{m\,s^{-1}}$;
  $(b)$~$\Uo = 10\,\mathrm{m\,s^{-1}}$.
  The unforced cases increase slowly for $x/h \gtrsim 20$; forced
  conditions are systematically reduced across the full downstream range,
  with both forcing periods collapsing.}
  \label{fig:Relam}
\end{figure}

The streamwise evolution of the turbulence scales for the unforced cases
(S-5, S-10) establishes the steady reference state. The integral length
scale $L$ grows monotonically along the jet axis
(figure~\ref{fig:Lint}$a,b$): it increases by a factor of approximately 3
from $x/h=5$ to $x/h=90$, consistent with the progressive coarsening of the
turbulent structure as the jet entrains ambient fluid and the large eddies
grow with the jet width. The Taylor-scale Reynolds number $\Relam$ decreases
sharply in the near field ($x/h\lesssim15$) as the potential core breaks
down and turbulence develops from the initial shear layers, then increases
slowly for $x/h\gtrsim20$ (figure~\ref{fig:Relam}). At the measurement
station $x/h=50$, the values are $\Relam\approx120$ and 212 for S-5 and
S-10, respectively, consistent with table~\ref{tab:conditions}.

The slow increase of $\Relam$ in the far field is a well-established
property of the self-similar planar jet \citep{Cafiero2019}. The normalised
dissipation coefficient $\Ceps$, defined in equation~\eqref{eq:ceps}, takes
values in the range $0.3$--$0.6$ across the far-field region for both
unforced conditions (figure~\ref{fig:Ceps}). Rather than approaching a
constant value, as expected from the standard dissipation scaling
$\eps\sim u_\mathrm{rms}^3/L$ consistent with Kolmogorov phenomenology,
the unforced data exhibit a systematic dependence
$\Ceps\propto\Relam^{-0.60\pm0.13}$, indistinguishable in slope from the
forced cases (\S\,\ref{sec:singlephase:noneq}). The standard scaling is
therefore not recovered over the comparatively limited streamwise range
investigated here, which nevertheless corresponds to distances of direct
relevance to many practical applications of turbulent jets. This behaviour
is consistent with previous measurements in canonical planar jets at
moderate $\Relam$ \citep{Cafiero2019}.

\subsection{Evolution of the turbulence scales under forcing}
\label{sec:singlephase:suppression}

\begin{figure}
  \includegraphics[width=\textwidth]{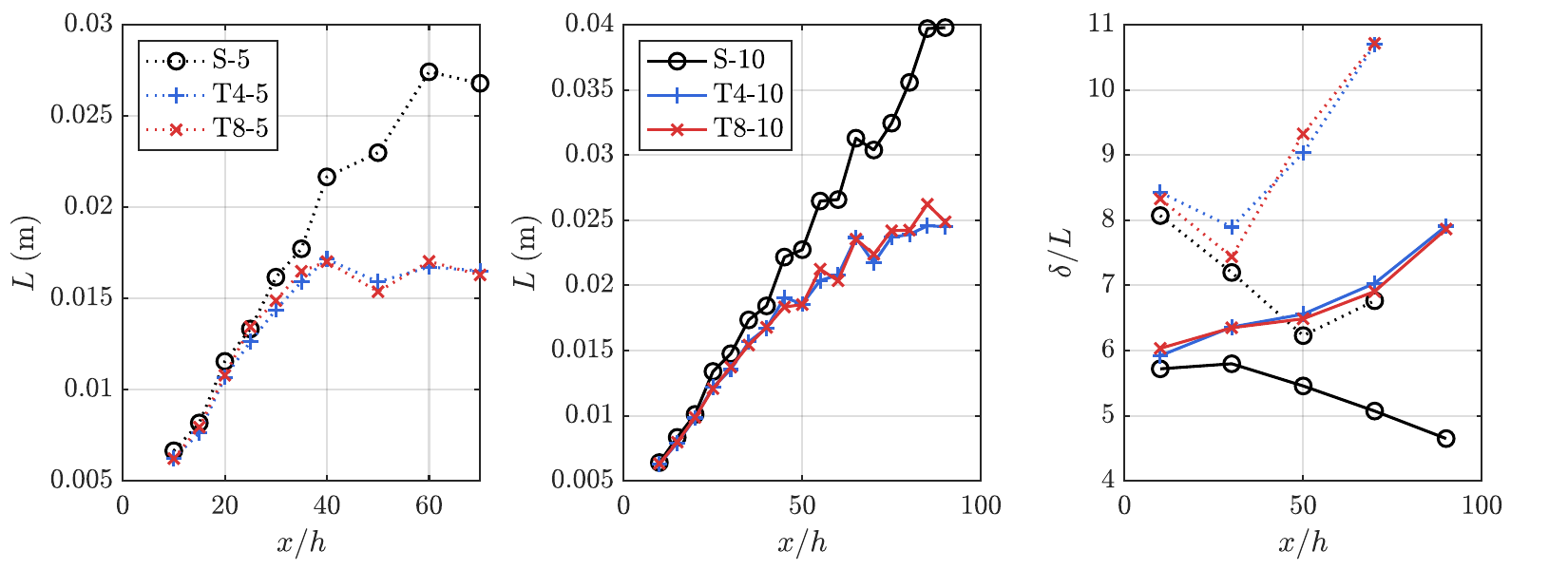}
  \caption{Integral scale and scale-separation ratio along the jet centreline.
  $(a,b)$~Integral scale $L(x/h)$ for $\Uo = 5$ and
  $10\,\mathrm{m\,s^{-1}}$, showing the arrest of $L$ growth under forcing
  and the $\approx40\%$ reduction at $x/h=50$.
  $(c)$~Ratio of integral scale to jet half-width, $L/\delta$, for all six
  conditions. The ratio approaches an approximately constant downstream
  value within each condition, while forcing shifts this value to a lower
  level, indicating that the proportionality $L\propto\delta$ is preserved
  with a forcing-dependent prefactor.}
  \label{fig:Lint}
\end{figure}

Periodic forcing produces clear and systematic modifications to the
turbulence scales across the full downstream development. At the measurement
station $x/h=50$, the Taylor-scale Reynolds number is reduced by
approximately $15\%$ for $\Uo=10\,\mathrm{m\,s^{-1}}$
(from 212 to 178--179) and by approximately $20\%$ for
$\Uo=5\,\mathrm{m\,s^{-1}}$ (from 120 to 93--97). The reduction of the
integral scale $L$ is more pronounced: the forced conditions reach values
approximately $40\%$ below the steady baseline at $x/h=50$ for
$\Uo=5\,\mathrm{m\,s^{-1}}$, with the difference persisting and increasing
downstream (figure~\ref{fig:Lint}$a,b$). Importantly, $L$ ceases to grow
beyond $x/h\approx40$--$50$ under forcing, whereas the unforced integral
scale continues increasing monotonically. Periodic modulation therefore
strongly modifies the downstream growth of the energy-containing length
scale despite leaving the time-averaged mean flow essentially unchanged.

These comparisons refer to the turbulent component isolated through the
triple decomposition described in \S\,\ref{sec:methods:td}, which separates
the deterministic periodic response from the stochastic velocity
fluctuations. Such a decomposition is particularly appropriate for
periodically forced turbulence and has been shown to provide a consistent
framework for characterising phase-dependent turbulence statistics and
dissipation under unsteady forcing \citep{Ahmad2026}.

Both forcing frequencies (T4 and T8) produce virtually identical values of
$\Relam$ and $L$ at all streamwise stations. Within the range investigated,
the modification of these quantities is therefore only weakly sensitive to
the precise forcing period. The same behaviour is observed at the two mean
exit velocities, indicating that the principal effect of the periodic
modulation is robust across the present operating conditions.

The changes in the relative organisation of the turbulence scales can be
summarised through two derived ratios. First, $L/\lambda$ provides a measure
of the separation between the integral and Taylor scales and is proportional
to $\Relam$ under the standard dissipation scaling consistent with
Kolmogorov phenomenology. Since both $L$ and $\Relam$ decrease under forcing,
$L/\lambda$ is also reduced, indicating a smaller separation between the
large and intermediate turbulence scales.

A particularly informative result is provided by the ratio $L/\delta$,
where $\delta$ is the jet half-width (figure~\ref{fig:Lint}$c$). The
proportionality $L\propto\delta$ is one of the underlying assumptions of
self-similar descriptions of free-shear flows, in which the characteristic
turbulence length scale grows in proportion to the width of the flow \citep{Townsend1976,cafiero2020length}.
Interestingly, this proportionality remains approximately satisfied for both
the steady and periodically forced jets: within each condition, $L/\delta$
approaches an approximately constant value over the downstream region.
Periodic forcing therefore does not destroy this aspect of self-similar
development, but changes its proportionality coefficient. The forced cases
settle onto a systematically lower value of $L/\delta$ than the corresponding
steady baselines, by as much as a factor of two, while retaining the same
approximately constant downstream scaling.

This distinction is important because $\delta$ itself is essentially
unaffected by forcing (\S\,\ref{sec:singlephase:meanflow}). The different
prefactors therefore arise from the reduced integral scale $L$, rather than
from a modification of the mean jet spreading. Thus, the mean-flow and
turbulence length scales remain proportional during their downstream
development, as required by the self-similar framework, but periodic forcing
changes the proportionality between them. The forcing consequently preserves
the form of the self-similar scaling while selecting a different relative
size of the energy-containing turbulent structures within the jet.

The separation between the forcing period and the intrinsic turbulence time
scales provides useful context for this behaviour. At $x/h=50$, the
integral-scale turnover time
$T_L=L/u_\mathrm{rms}\approx0.07$--$0.15\,\mathrm{s}$ is much shorter than
the forcing period. The turbulence therefore has sufficient time to adjust
substantially within each forcing cycle, but the repeated modulation prevents
the flow from experiencing the same downstream evolution as under steady
conditions. The resulting time-averaged effect is a systematic reduction of
$L$ and $\Relam$, while the large-scale mean-flow development remains
essentially unchanged.

\subsection{Common $\Ceps(\Relam)$ scaling across all conditions}
\label{sec:singlephase:noneq}

\begin{figure}
  \includegraphics[width=\textwidth]{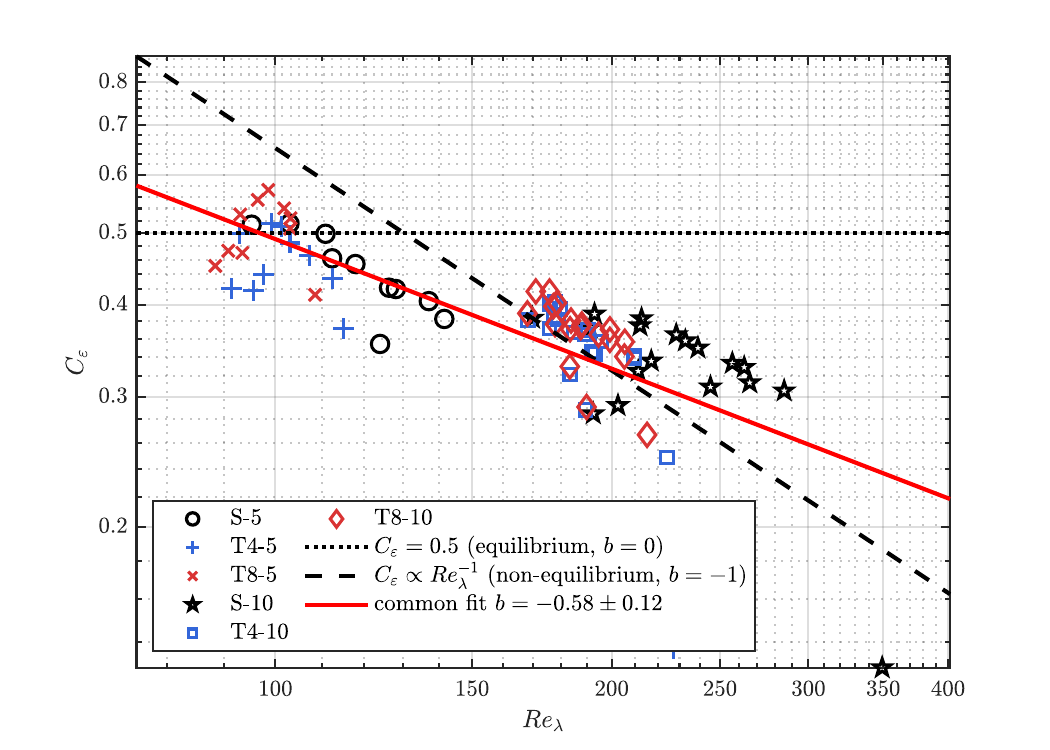}
  \caption{Normalised dissipation coefficient $\Ceps$ as a function of
  Taylor-scale Reynolds number $\Relam$, using all six conditions and
  every resolved streamwise station as individual data points
  ($n=87$). Solid line: common log--log fit
  $\Ceps=A\,\Relam^{b}$ with
  $b=-0.58\pm0.08$ (95\% CI), $\log_{10}A=+0.86$, and residual standard
  deviation $\sigma(\log_{10}\Ceps)=0.08$. Reference dashed lines:
  $\Ceps=0.5$ (constant-$\Ceps$ scaling, $b=0$) and
  $\Ceps\propto\Relam^{-1}$ (non-equilibrium scaling, $b=-1$).
  All six conditions lie on the common branch; forcing modifies the range
  occupied along the curve, reducing both $\Relam$ and $\Ceps$, rather
  than producing a distinct scaling relation.}
  \label{fig:Ceps}
\end{figure}

A complementary description of the downstream turbulence evolution is
obtained by considering the normalised dissipation coefficient $\Ceps$ as a
function of $\Relam$. Figure~\ref{fig:Ceps} shows the measurements from all
resolved streamwise stations for the six experimental conditions. The two
unforced cases (S-5, S-10) and the four forced cases (T4-5, T8-5, T4-10,
T8-10) collapse onto a common power-law branch,
\begin{equation}
  \Ceps = A\,\Relam^{b}, \qquad
  b = -0.58 \pm 0.08,\quad \log_{10} A = +0.86,
  \label{eq:noneq_scaling}
\end{equation}
with a residual standard deviation $\sigma(\log_{10}\Ceps)=0.08$ across
the full range $\Relam\in[80,230]$ (87 $(\text{case},x/h)$ data points,
pooling all six conditions over every streamwise station that passes the
far-field convergence criteria of \S\,\ref{sec:methods}; near-field stations
within the potential core are excluded).

Fitting the steady cases (S-5, S-10) and forced cases separately gives
$b_\mathrm{steady}=-0.60\pm0.13$ and
$b_\mathrm{forced}=-0.58\pm0.08$, respectively, which are indistinguishable
within their 95\% confidence intervals. The unforced and forced populations
therefore do not separate in the $(\Relam,\Ceps)$ plane. In this respect,
the common $\Ceps(\Relam)$ relationship belongs to the same set of
time-averaged properties as the mean-flow evolution discussed in
\S\,\ref{sec:singlephase:meanflow}: periodic forcing modifies the values of
the turbulence quantities attained at a given downstream location without
producing a detectable change in the functional relation between $\Ceps$
and $\Relam$. Both $\Relam$ and $\Ceps$ are reduced under forcing
(\S\,\ref{sec:singlephase:suppression}), so the forced conditions occupy a
different range along the same branch rather than a distinct scaling regime.

The fitted exponent $b\approx-0.6$ lies between the constant-$\Ceps$
behaviour ($b=0$) associated with the standard dissipation scaling
$\eps\sim u_\mathrm{rms}^3/L$ and the $b=-1$ scaling proposed for
non-equilibrium turbulence in the large $Re_\lambda$ regime by \citet{Vassilicos2015}. The present
measurements therefore occupy an intermediate range in which the standard
scaling consistent with Kolmogorov phenomenology is not recovered over the
streamwise distances investigated. Importantly, this behaviour is already
observed in the unforced jet; periodic forcing does not produce a different
dissipation scaling, but rather changes the range of $\Relam$ and $\Ceps$
sampled along the same empirical relation. Intermediate exponents between
these limiting behaviours have also been reported in other canonical
turbulent flows \citep{Vassilicos2015,Goto2016}.

Within the framework proposed by \citet{Cafiero2019}, in which
$\Ceps\sim(Re_G/Re_\delta)^m$, the measured exponent corresponds to
$m\approx0.58\pm0.08$. This value lies between the standard
constant-$\Ceps$ case ($m=0$, $b=0$) and the non-equilibrium limit
($m=1$, $b=-1$). The associated mean-flow decay exponent,
$a=(m+1)/(4m+2)\approx0.37$, lies between the corresponding limiting values
$1/2$ and $1/3$ and remains compatible with the mean-flow measurements of
\S\,\ref{sec:singlephase:meanflow}, for which the finite streamwise range
and the uncertainty associated with the virtual origin make a direct
discrimination between these relatively close exponents difficult.

Comparable intermediate behaviours have been reported in spatially
developing turbulent flows, including wakes
\citep{Dairay2015,ortiz2021high}, decaying grid turbulence
\citep{Valente2012,Hearst2015}, and planar jets \citep{Cafiero2019}.
Non-equilibrium dissipation scalings have also been observed in direct
numerical simulations (DNS) of turbulent wakes \citep{Dairay2016}
and turbulent fronts \citep{Zecchetto2023}.
More recently, DNS studies of unsteady turbulence have highlighted
the role of large-scale dynamics and forcing transients in modifying
the relationship between the normalised dissipation and the
Taylor-scale Reynolds number \citep{Zapata2024,Ghira2026}.
The present measurements show that, over the streamwise range
considered here, periodic forcing does not alter this time-averaged
$\Ceps(\Relam)$ scaling despite producing substantial changes in
$L$ and $\Relam$. This distinction is important: the forcing modifies
the downstream turbulence state while the functional relationship
between the normalised dissipation and the Taylor-scale Reynolds
number remains approximately invariant.

Temporal forcing can nevertheless produce instantaneous behaviour that is
not apparent from time-averaged quantities. In forced grid turbulence,
\citet{Ahmad2026} reported phase-dependent hysteresis in the
$(Re_\lambda,C_\varepsilon)$ plane, demonstrating that identical values of
one turbulence parameter need not correspond to identical instantaneous
states under unsteady conditions. The present time-averaged
$\Ceps(\Relam)$ measurements do not by themselves establish such a
hysteretic response. Indeed, as shown in
\S\,\ref{sec:singlephase:phase}, $\Relam(\vph)$ varies only weakly within
the forcing cycle. They nevertheless motivate examining the turbulence
quantities directly as functions of forcing phase.

The ratio of the characteristic time scales further supports this
phase-resolved analysis. At $x/h=50$,
$T_L=L/u_\mathrm{rms}\approx0.07$--$0.15\,\mathrm{s}$, yielding
$T_L/T_f\approx10^{-2}$. The integral-scale turnover time is therefore
approximately two orders of magnitude shorter than the forcing period.
This strong separation of time scales suggests that the turbulence can
respond appreciably during each cycle, even though the forcing remains slow
relative to the characteristic turbulent motion. Similar sensitivity of
dissipation statistics to continuously varying large-scale conditions has
been discussed by \citet{Goto2016}. The phase-resolved measurements below
allow this unsteadiness to be examined directly, without requiring the
time-averaged $\Ceps(\Relam)$ relation itself to be interpreted as evidence
of a distinct dynamical regime.

\subsection{Phase-resolved turbulence dynamics within the forcing cycle}
\label{sec:singlephase:phase}

The time-averaged results described above do not reveal how the turbulence
varies within an individual forcing cycle. The triple decomposition
(\S\,\ref{sec:methods:td}) allows the turbulence statistics to be tracked as
a function of forcing phase $\vph$, providing a direct characterisation of
the instantaneous unsteadiness.

\begin{figure}
 \includegraphics[width=\textwidth, trim={1cm 7.4cm 1cm 7.4cm},clip]
 {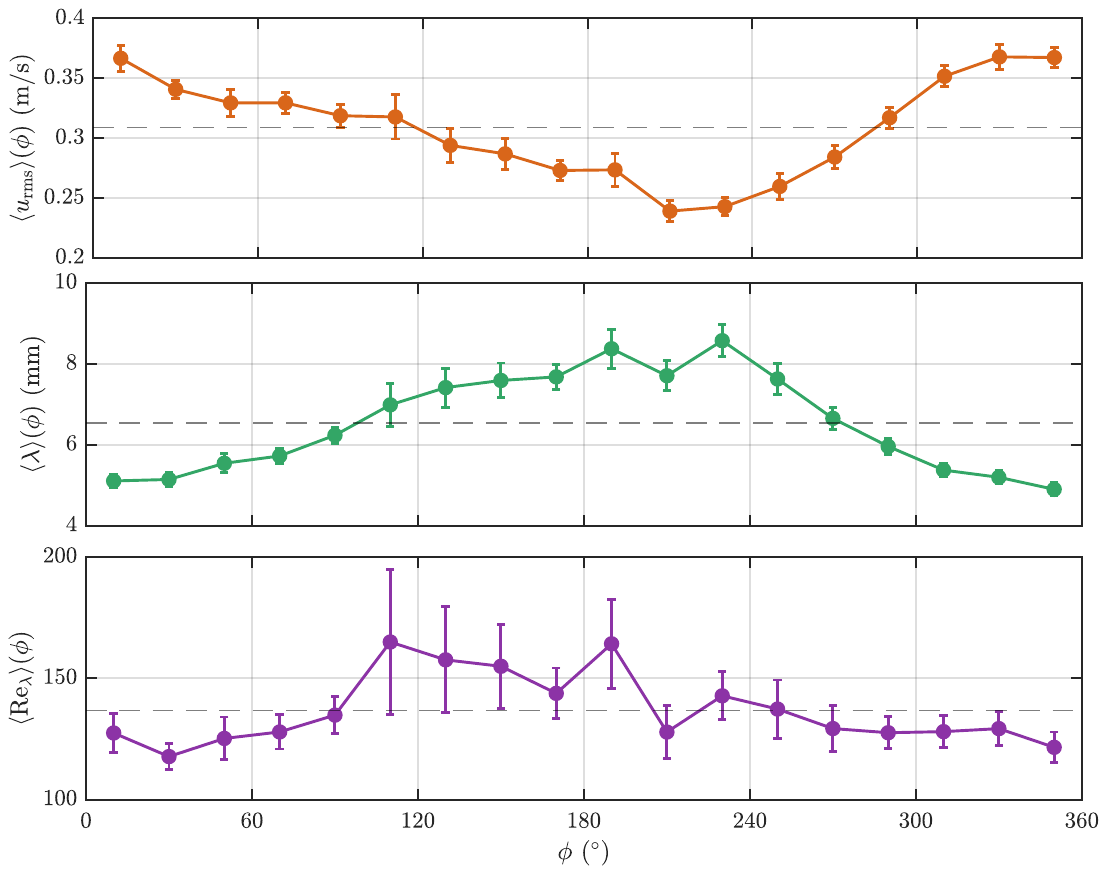}
  \caption{Phase-averaged single-phase turbulence statistics at $x/h=50$
  for the T8-5 case ($\Uo=5\,\mathrm{m\,s^{-1}}$, $T_f=8\,\mathrm{s}$).
  Three panels, from top to bottom:
  $\langle u_\mathrm{rms}\rangle(\vph)$ (orange),
  $\langle\lambda\rangle(\vph)$ (green), and
  $\langle\Relam\rangle(\vph)$ (purple).
  Horizontal dashed lines indicate cycle means.
  $u_\mathrm{rms}(\vph)$ and $\lambda(\vph)$ oscillate with comparable
  amplitudes and approximately in anti-phase, while their product and hence
  $\Relam(\vph)$ vary only weakly ($\pm10\%$).}
  \label{fig:phasehwa}
\end{figure}

Figure~\ref{fig:phasehwa} shows the phase-averaged profiles
$\langle u_\mathrm{rms}\rangle(\vph)$,
$\langle\lambda\rangle(\vph)$, and
$\langle\Relam\rangle(\vph)$ at $x/h=50$ for the T8-5 case. The same
qualitative behaviour is observed in the T4-5 case and in both
$\Uo=10\,\mathrm{m\,s^{-1}}$ forced conditions; T8-5 is presented because
the longer forcing period provides the largest number of phase-resolved
samples per bin and the clearest statistical convergence. The oscillations
are not artefacts of the averaging procedure, as they are also visible
cycle by cycle in the instantaneous records.

The turbulent velocity fluctuation
$\langle u_\mathrm{rms}\rangle(\vph)$ responds clearly to the forcing. It
reaches a maximum near $\vph=0^\circ/360^\circ$, corresponding to the end
of the accelerating half-cycle, at approximately
$0.34\,\mathrm{m\,s^{-1}}$, and decreases to approximately
$0.21\,\mathrm{m\,s^{-1}}$ near
$\vph\approx200^\circ$--$210^\circ$. This represents a peak-to-peak
variation of approximately $45\%$ about the cycle mean. The Taylor
microscale $\langle\lambda\rangle(\vph)$ exhibits a comparable relative
variation but evolves in the opposite sense: $\lambda$ is smallest
(approximately $4$--$4.5\,\mathrm{mm}$) when $u_\mathrm{rms}$ is largest
and reaches approximately $7\,\mathrm{mm}$ near the minimum of
$u_\mathrm{rms}$. The two quantities are therefore approximately in
anti-phase.

The consequence of this anti-phase evolution is visible in the third panel.
The Taylor-scale Reynolds number
\begin{equation}
\langle\Relam\rangle(\vph)
=
\frac{\langle u_\mathrm{rms}\rangle(\vph)
      \langle\lambda\rangle(\vph)}{\nu}
\end{equation}
remains nearly constant throughout the forcing cycle, with variations not
exceeding approximately $\pm10\%$ about its mean value. The anti-correlation
between $u_\mathrm{rms}$ and $\lambda$ is strong but not exact: the change in
$\lambda$ nearly compensates that of $u_\mathrm{rms}$, so that their product
is only weakly modulated rather than strictly conserved.

This behaviour highlights an important limitation of using $\Relam$ alone
to characterise the instantaneous turbulence state. Although
$\Relam(\vph)$ varies only weakly, its constituent quantities undergo
substantial and systematic phase-dependent changes. Writing the measured
relationship as
$\lambda\propto u_\mathrm{rms}^{-\alpha}$ gives
$\alpha\approx0.83$, so that
$\Relam\propto u_\mathrm{rms}^{1-\alpha}
=u_\mathrm{rms}^{0.17}$, consistent with the observed weak variation.
Using
$\lambda=(15\nu u_\mathrm{rms}^2/\eps)^{1/2}$ then gives
$\eps\propto u_\mathrm{rms}^{2(1+\alpha)}
\approx u_\mathrm{rms}^{3.7}$.

For comparison, the standard dissipation scaling
$\eps\sim u_\mathrm{rms}^3/L$, consistent with Kolmogorov phenomenology,
would imply a weaker dependence under otherwise comparable conditions.
The present phase-resolved measurements therefore show that substantial
variations in the dissipation and turbulence scales can occur while
$\Relam$ remains nearly unchanged. Such departures from the standard
dissipation scaling under unsteady conditions are consistent with previous
observations in temporally evolving turbulence
\citep{Obligado2019,Zheng2023}.

The weak modulation of $\Relam(\vph)$ should therefore not be interpreted
as an absence of an unsteady response. Rather, it results from the opposing
variations of $u_\mathrm{rms}$ and $\lambda$. At the same time, the
time-averaged value of $\Relam$ remains systematically lower in the forced
cases than in the corresponding unforced baselines
(\S\,\ref{sec:singlephase:suppression}). The forcing thus affects the
turbulence on two distinct levels: it modifies the downstream,
time-averaged evolution of the turbulence scales and also produces a
reproducible phase dependence within each forcing cycle.

This distinction becomes particularly important for the dispersed phase.
Despite the weak variation of $\Relam(\vph)$, the Kolmogorov time scale
\begin{equation}
\langle\taueta\rangle(\vph)
=
\left(
\frac{\nu}{\langle\eps\rangle(\vph)}
\right)^{1/2}
\end{equation}
varies substantially through the cycle. Since
$\eps\propto u_\mathrm{rms}^2/\lambda^2$ and
$\lambda\propto u_\mathrm{rms}^{-\alpha}$ with
$\alpha\approx0.83$, the measured scaling gives
$\taueta\propto u_\mathrm{rms}^{-(1+\alpha)}
\approx u_\mathrm{rms}^{-1.85}$. The Kolmogorov time scale therefore varies
in the opposite sense to $u_\mathrm{rms}$ and with a comparatively strong
relative amplitude.

As a consequence, the effective particle Stokes number
$\St(\vph)=\taup/\taueta(\vph)$ also varies throughout the forcing cycle even
though $\Relam(\vph)$ remains nearly constant. Thus, particles experience
different instantaneous small-scale conditions at phases characterised by
similar values of $\Relam$. This observation provides the direct
fluid-mechanical link to the particle measurements discussed in
\S\,\ref{sec:twophase}, where the phase-resolved clustering response is
compared with the evolution of these instantaneous turbulence quantities.

\section{Results: particle-laden flow}
\label{sec:twophase}

\subsection{Preferential concentration in the steady reference flow}
\label{sec:twophase:steady}

\begin{figure}
  \centerline{\includegraphics[width=0.70\textwidth,
    trim={1cm 7.4cm 1cm 7.4cm},clip]%
    {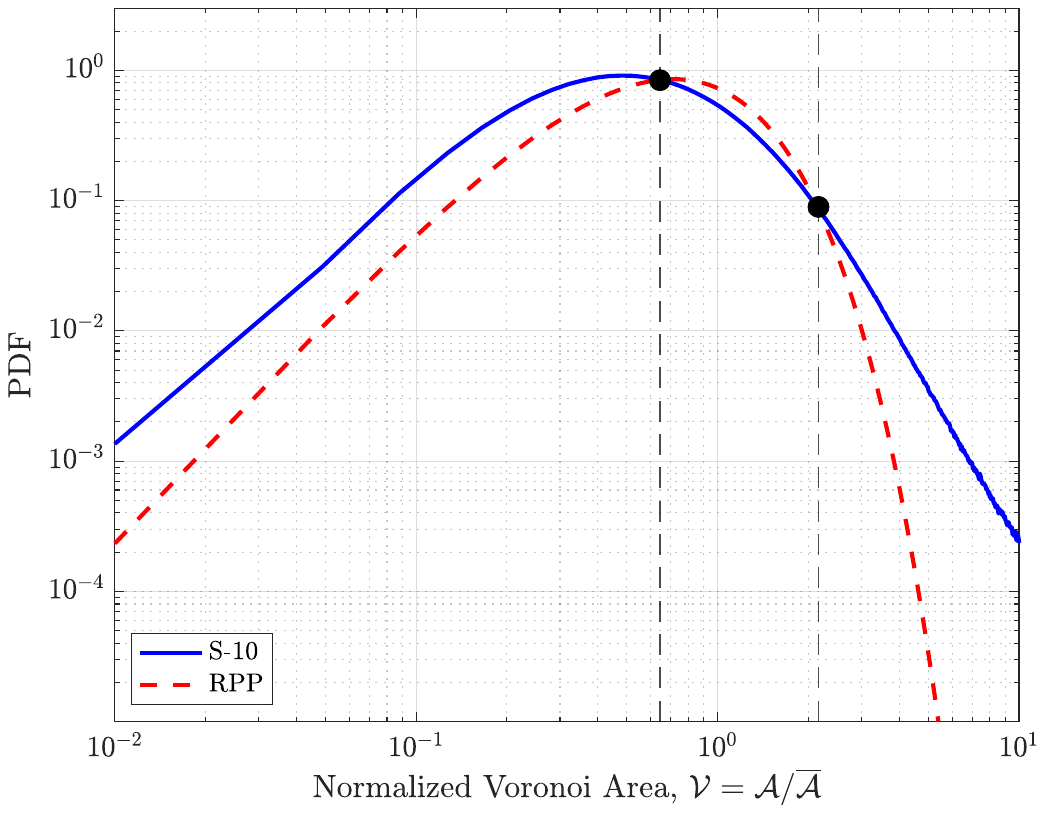}}
  \caption{Probability density function of the normalised Voronoi cell area
  $\mathcal{A}/\overline{\mathcal{A}}$ for the S-10 case
  ($\Uo=10\,\mathrm{m\,s^{-1}}$, unforced), compared with the random
  Poisson process (RPP, dashed red line). Filled circles mark the cluster
  threshold ($\mathcal{A}/\overline{\mathcal{A}}=0.62$) and void threshold
  ($\mathcal{A}/\overline{\mathcal{A}}=2.14$) at the intersections of the
  two distributions. The experimental distribution exceeds the RPP in both
  tails, confirming significant preferential concentration.}
  \label{fig:voronoi_pdf}
\end{figure}

Before examining the response to periodic forcing, we first establish the
clustering state of the droplets in the unforced reference conditions.
Figure~\ref{fig:voronoi_pdf} shows the probability density function of the
normalised Voronoi cell area
$\mathcal{A}/\overline{\mathcal{A}}$ for S-10, together with the
corresponding random Poisson process (RPP) reference introduced in
\S\,\ref{sec:voronoi}; S-5 exhibits qualitatively similar behaviour.
We remark that the imaging region is entirely contained within the particle-laden core of
the jet and does not intersect the jet edge or a turbulent/non-turbulent
interface, thereby avoiding contributions associated with intermittency at
the jet boundary.

The experimental distribution departs clearly from the RPP reference, with
enhanced probabilities of both small and large Voronoi cells. Following the
interpretation introduced in \S\,\ref{sec:voronoi}, this confirms that
preferential concentration is already well established in the unforced jet
and provides the reference state against which the effects of periodic
forcing are assessed. Such spatial organisation is consistent with the
preferential sampling of the carrier flow by inertial particles
\citep{Maxey1987,Squires1991,Eaton1994}.

For the measured droplet size distribution (\S\,\ref{sec:setup:imaging}),
the droplets operate at Stokes numbers of order unity at $x/h=50$: based on
the Sauter diameter, $\St\approx0.56$ at $\Uo=5\,\mathrm{m\,s^{-1}}$ and
$\approx2.0$ at $\Uo=10\,\mathrm{m\,s^{-1}}$. They therefore lie on the
small-inertia side of the preferential-concentration maximum at
$\St=O(1)$ for $\Uo=5$, and on its large-inertia side for $\Uo=10$. Voronoi measurements of droplets at Stokes numbers of order unity in wind-tunnel
turbulence indicate, however, that their preferential concentration is not
confined to the dissipative scales \citep{Sumbekova2017}. The role of the
instantaneous Stokes number is considered more directly in
\S\,\ref{sec:twophase:phase}.

The intersections between the experimental and RPP distributions, defined
according to the criterion of \citet{Monchaux2010}, occur at
$\mathcal{A}/\overline{\mathcal{A}}=0.62$ and $2.14$ for clusters and
voids, respectively. In the following, the overall clustering intensity is
quantified primarily through the standard deviation $\sigV$, as defined in
\S\,\ref{sec:voronoi}.

For the unforced conditions S-5 and S-10,
$\langle\sigV\rangle/\sigRPP\approx2.0$ and $1.6$, respectively. These
values establish the steady reference levels against which the forced cases
are compared. Importantly, the stronger clustering at the lower exit
velocity is already present without forcing. The two reference conditions
differ not only in $\Relam$ but also in their Kolmogorov time scale and
therefore in the particle Stokes number. The difference between S-5 and
S-10 should consequently not be interpreted as a dependence on $\Relam$
alone.
At the lower exit velocity, the longer $\taueta$ gives a smaller particle
Stokes number ($\St\approx0.56$ against $\approx2.0$ at
$\Uo=10\,\mathrm{m\,s^{-1}}$), closer to the range of strongest
preferential concentration in statistically stationary turbulence
\citep{Bec2007,Monchaux2012}. The stronger clustering at
$\Uo=5\,\mathrm{m\,s^{-1}}$ can therefore not be attributed to the Stokes
number alone; the two conditions also differ in the droplet volume
fraction, which is twice as large at the lower exit velocity
(\S\,\ref{sec:setup:imaging}). This ordering also provides the
baseline for the forced conditions examined below.

\subsection{Effect of periodic forcing on time-averaged clustering}
\label{sec:twophase:tavg}

\begin{figure}
  \centerline{\includegraphics[width=\textwidth]%
    {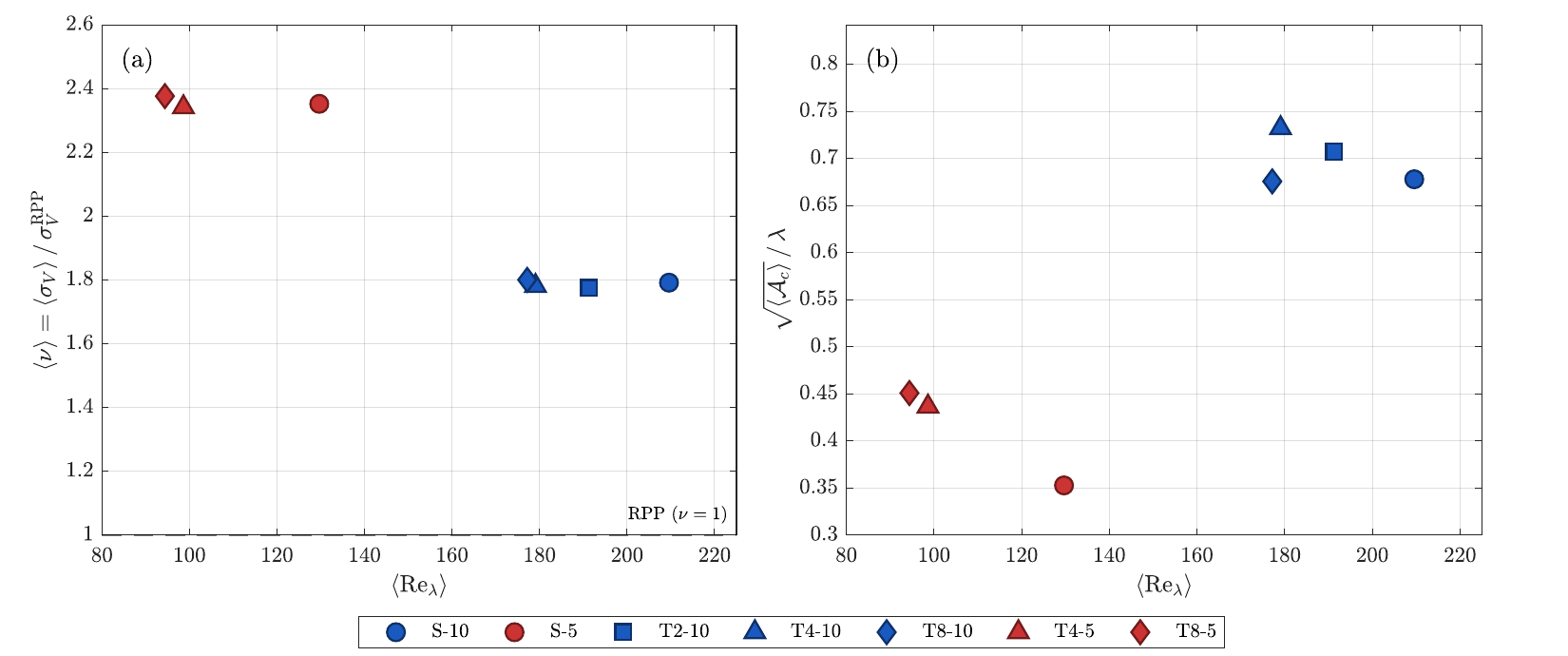}}
  \caption{Two clustering measures as functions of the time-averaged
  Taylor-scale Reynolds number $\langle\Relam\rangle$ for all measured
  two-phase conditions.
  $(a)$~Time-averaged Voronoi clustering intensity
  $\langle\sigV\rangle/\sigRPP$.
  $(b)$~Taylor-normalised mean cluster length
  $\sqrt{\langle\mathcal{A}_c\rangle}/\lambda$, where
  $\langle\mathcal{A}_c\rangle$ is the mean physical area of the Voronoi
  clusters and $\lambda$ is the Taylor microscale from the hot-wire
  measurements at $\xh=50$. Colours denote exit velocity
  ($\Uo=5\,\mathrm{m\,s^{-1}}$ red;
  $\Uo=10\,\mathrm{m\,s^{-1}}$ blue); marker shapes denote the experimental
  conditions. The clustering intensity is higher for the lower-velocity
  group, whereas the Taylor-normalised cluster length exhibits the opposite
  ordering.}
  \label{fig:clusterarea}
\end{figure}

Figure~\ref{fig:clusterarea} reports two complementary time-averaged
measures of the droplet clustering for all experimental conditions as
functions of the time-averaged Taylor-scale Reynolds number
$\langle\Relam\rangle$. Panel~(a) shows the normalised Voronoi standard
deviation $\langle\sigV\rangle/\sigRPP$, while panel~(b) shows the
Taylor-normalised mean cluster length
$\sqrt{\langle\mathcal{A}_c\rangle}/\lambda$, where
$\langle\mathcal{A}_c\rangle$ is the mean physical area of the identified
Voronoi clusters.

The forced conditions form two well-separated groups. At
$\Uo=5\,\mathrm{m\,s^{-1}}$, T4-5 and T8-5 have
$\langle\Relam\rangle\approx94$--$99$ and
$\langle\sigV\rangle/\sigRPP\approx2.3$--$2.4$. At
$\Uo=10\,\mathrm{m\,s^{-1}}$, T4-10 and T8-10 have
$\langle\Relam\rangle\approx177$--$179$ and
$\langle\sigV\rangle/\sigRPP\approx1.8$.

Three features are noteworthy. First, within each exit-velocity group, the
two forcing periods produce very similar values of
$\langle\sigV\rangle/\sigRPP$. This mirrors the weak sensitivity of the
single-phase turbulence statistics to the precise forcing period observed in
\S\,\ref{sec:singlephase:suppression}. Within the range investigated,
changing $T_f$ therefore has little influence on the time-averaged
clustering state.

Second, the lower-$\Relam$ group exhibits stronger clustering than the
higher-$\Relam$ group. This ordering should not be interpreted as a direct
Reynolds-number dependence because the two groups also operate at different
particle Stokes numbers and droplet volume fractions. Preferential concentration in statistically
stationary turbulence depends strongly on particle inertia
\citep{Bec2007,Monchaux2010}, and the differences in $\taueta$ between the
two exit velocities imply different effective values of $\St$. The present
data therefore already indicate that $\Relam$ alone is insufficient to
compare the clustering state across conditions with different small-scale
time scales. The substantial values measured here,
$\langle\sigV\rangle/\sigRPP\approx1.8$--$2.4$, confirm that preferential
concentration remains pronounced under the present conditions, consistent
with previous observations of significant inertial-particle clustering in
turbulent flows \citep{Sumbekova2017,Obligado2014}.

Third, and more importantly for the present study, periodic forcing produces
a systematic increase in the time-averaged clustering intensity relative to
the corresponding unforced condition. At $\Uo=5\,\mathrm{m\,s^{-1}}$,
$\langle\sigV\rangle/\sigRPP$ increases from approximately $2.0$ in S-5
to $2.3$--$2.4$ in T4-5 and T8-5. At
$\Uo=10\,\mathrm{m\,s^{-1}}$, it increases from approximately $1.6$ in
S-10 to $1.8$ in the forced cases. Periodic forcing therefore affects not
only the instantaneous clustering dynamics, as shown below, but also the
mean preferential-concentration level.

The cluster geometry in figure~\ref{fig:clusterarea}(b) contains
complementary information. The Taylor-normalised mean cluster length
increases from
$\sqrt{\langle\mathcal{A}_c\rangle}/\lambda\approx0.35$--$0.45$ at
$\Uo=5\,\mathrm{m\,s^{-1}}$ to approximately $0.68$--$0.73$ at
$\Uo=10\,\mathrm{m\,s^{-1}}$, with the two forcing periods again giving
similar values within each group. Thus, the conditions exhibiting the
strongest concentration inhomogeneity are not those with the largest
clusters relative to $\lambda$. The Voronoi standard deviation and the
cluster length therefore quantify distinct aspects of preferential
concentration: the former measures its intensity, whereas the latter
characterises the spatial extent of the clustered regions.

The ordering of the time-averaged data follows the small-scale time scales
experienced by the particles: clustering is stronger where $\taueta$ is
longer. The Stokes number is
$\St=\taup/\taueta$, so variations in $\taueta$ directly change the
effective particle inertia even though the particle properties remain
fixed. The two exit velocities produce different values of $\taueta$ and
therefore place the droplets at different positions along the
clustering-versus-$\St$ dependence. Moreover, as shown in
\S\,\ref{sec:singlephase:phase}, periodic forcing produces substantial
variations of the dissipation and $\taueta$ even when $\Relam$ varies only
weakly. The resulting particle response is therefore more naturally
interpreted through the instantaneous small-scale time scales than through
$\Relam$ alone. This point is examined directly using the phase-resolved
measurements in the following sections.

\subsection{Phase-resolved clustering response}
\label{sec:twophase:raw}

The time-averaged measurements demonstrate that forcing modifies the mean
clustering level, but do not reveal how the particle distribution evolves
within a forcing cycle. To resolve this response, each imaging frame is
assigned the forcing phase
$\vph\in[0^\circ,360^\circ)$ using the synchronised forcing signal, and
$\sigV/\sigRPP$ is averaged within the same $N_\vph=18$ phase intervals of
width $20^\circ$ employed for the single-phase measurements. Each bin
contains $\mathcal{O}(500)$ imaging frames, providing sufficient statistical
convergence despite the substantial frame-to-frame fluctuations of the
instantaneous Voronoi statistic.

\begin{figure}
  \centerline{\includegraphics[width=\textwidth]%
    {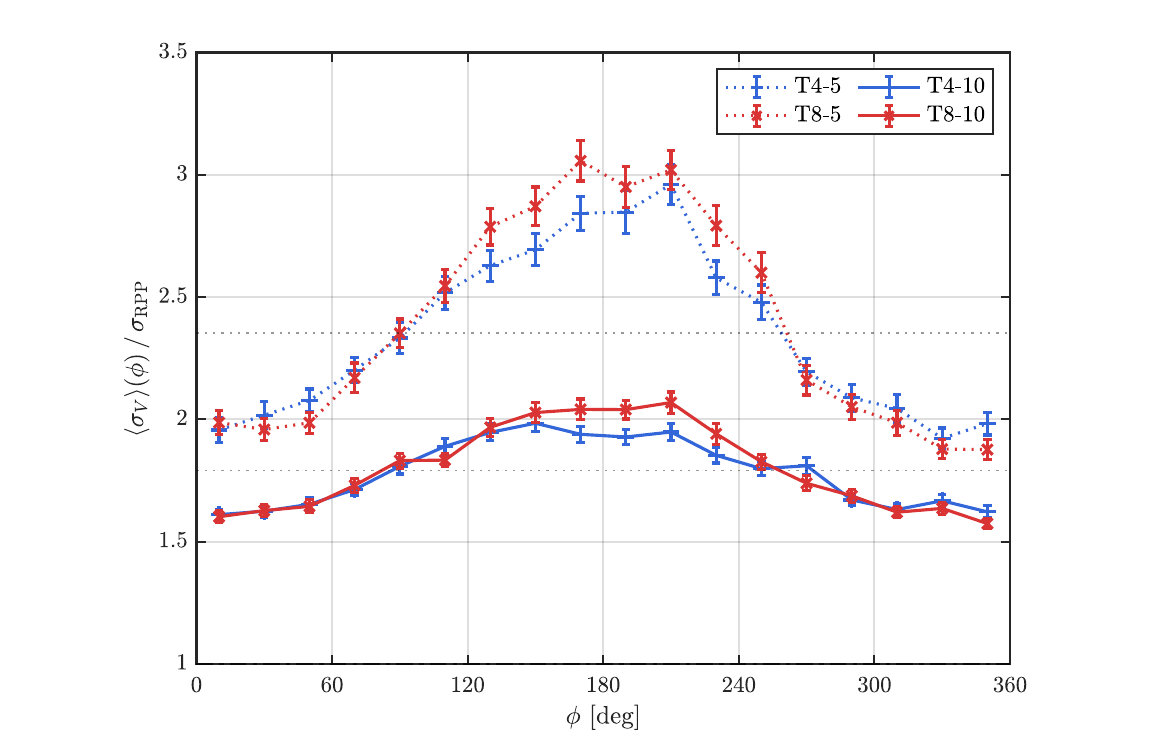}}
  \caption{Phase-averaged normalised clustering intensity
  $\langle\sigV\rangle(\vph)/\sigRPP$ as a function of forcing phase for
  the four periodically forced conditions. Dotted curves denote the
  $\Uo=5\,\mathrm{m\,s^{-1}}$ cases and solid curves the
  $\Uo=10\,\mathrm{m\,s^{-1}}$ cases; symbols distinguish the two forcing
  periods. Error bars indicate the standard error within each phase bin.
  The clustering intensity exhibits a clear phase-periodic modulation in
  all four conditions, with maxima occurring during the second half of the
  forcing cycle. Horizontal reference levels indicate the corresponding
  cycle means.}
  \label{fig:sigma_raw}
\end{figure}

Figure~\ref{fig:sigma_raw} shows the phase-resolved clustering intensity
$\langle\sigV\rangle(\vph)/\sigRPP$ for all four forced conditions. A clear
single-humped modulation is observed in every case. At
$\Uo=5\,\mathrm{m\,s^{-1}}$, the peak-to-peak variation amounts to
approximately $44\%$ and $49\%$ of the cycle mean for T4-5 and T8-5,
respectively. At $\Uo=10\,\mathrm{m\,s^{-1}}$, the corresponding
modulations are approximately $21\%$ and $27\%$ for T4-10 and T8-10.
Thus, the particle field exhibits a reproducible response at the forcing
frequency despite the relatively weak and slow modulation imposed at the
jet inlet.

The maxima reach
$\langle\sigV\rangle/\sigRPP\approx2.96$ for T4-5 and $3.06$ for T8-5,
and approximately $1.98$ and $2.07$ for T4-10 and T8-10, respectively.
The maxima therefore exceed the corresponding steady reference values,
whereas the cycle minima approach or fall slightly below them. The increase
in the time-averaged clustering reported in
\S\,\ref{sec:twophase:tavg} consequently results from a strongly
asymmetric modulation over the cycle, with phases of enhanced clustering
reaching farther above the steady state than the minima extend below it.

The two forcing periods again display very similar behaviour within each
exit-velocity group, although the phase of the maximum is not identical.
The clustering peaks occur at
$\vph\approx210^\circ$ (T4-5),
$170^\circ$ (T8-5),
$150^\circ$ (T4-10), and
$210^\circ$ (T8-10). In all cases, the maximum therefore occurs during the
second half of the forcing cycle, close to the phase at which the
single-phase measurements indicate the weakest small-scale turbulence.

Comparison with the phase-resolved hot-wire measurements of
\S\,\ref{sec:singlephase:phase} shows that
$\langle u_\mathrm{rms}\rangle(\vph)$ and
$\langle\eps\rangle(\vph)$ reach their minima approximately between
$\vph=190^\circ$ and $230^\circ$, while
$\langle\taueta\rangle(\vph)$ reaches its maximum over the same portion of
the cycle. The clustering maxima therefore occur close to the phases of
largest $\taueta$ and smallest dissipation. Across the four conditions, the
offset between the maximum of $\langle\sigV\rangle$ and that of
$\langle\taueta\rangle$ lies within approximately one to three phase bins.
The phase agreement is therefore systematic, although the present resolution
does not support a claim of exact zero-lag locking.

The same qualitative relationship is observed in the correlation
statistics. The correlation between $\sigV$ and $\eps$ is negative for all
four forced conditions: stronger preferential concentration is consistently
associated with lower instantaneous dissipation. The correlation strength
varies between conditions, however, with three cases displaying moderate
values $|r|\approx0.5$--$0.7$ and the strongest response occurring for
T8-5. The robust result is therefore the sign and phase organisation of the
response rather than an exact universal correlation coefficient.

This behaviour is consistent with the numerical results of
\citet{Zapata2024}, who showed that temporal variations of the carrier
turbulence can produce substantial modulation of inertial-particle
clustering. The present measurements extend this observation to a spatially
developing free-shear flow and show experimentally that the clustering
response remains clearly phase organised even when the Taylor-scale Reynolds
number itself varies only weakly.

\subsection{Relation between clustering and the instantaneous turbulence state}
\label{sec:twophase:phase}

\begin{figure}
  \centerline{\includegraphics[width=\textwidth]%
    {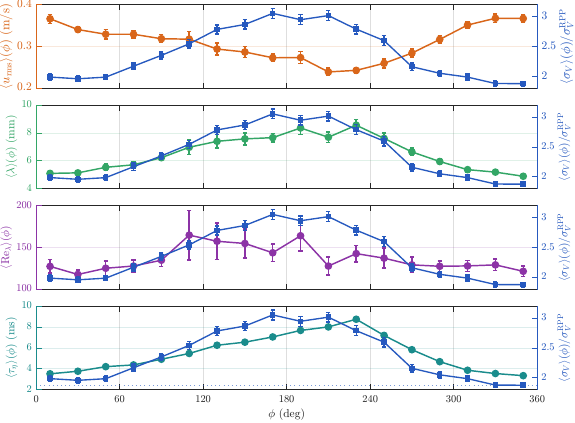}}
  \caption{Phase-averaged comparison of carrier-flow turbulence and
  clustering intensity for the T8-5 case
  ($\Uo=5\,\mathrm{m\,s^{-1}}$, $T_f=8\,\mathrm{s}$).
  Left axes, from top to bottom:
  $\langle u_\mathrm{rms}\rangle(\vph)$,
  $\langle\lambda\rangle(\vph)$,
  $\langle\Relam\rangle(\vph)$, and
  $\langle\taueta\rangle(\vph)$.
  Right axes in all panels:
  $\langle\sigV\rangle(\vph)/\sigRPP$.
  Horizontal dashed lines indicate cycle means. The clustering varies
  approximately in anti-phase with $u_\mathrm{rms}$, in phase with
  $\lambda$ and $\taueta$, and exhibits substantial modulation while
  $\Relam$ remains nearly constant.}
  \label{fig:phasecluster}
\end{figure}

The relation between the clustering response and the carrier turbulence is
examined more directly in figure~\ref{fig:phasecluster}. For the T8-5 case,
the phase-resolved clustering intensity
$\langle\sigV\rangle(\vph)/\sigRPP$ is compared with the single-phase
quantities introduced in \S\,\ref{sec:singlephase:phase}:
$\langle u_\mathrm{rms}\rangle(\vph)$,
$\langle\lambda\rangle(\vph)$,
$\langle\Relam\rangle(\vph)$, and
$\langle\taueta\rangle(\vph)$. T8-5 is shown because it exhibits the
strongest phase-resolved response, while the same qualitative ordering is
observed in the remaining forced conditions.

The first panel shows that the clustering varies approximately in anti-phase
with the turbulent velocity fluctuation. Values of
$\langle\sigV\rangle/\sigRPP$ are lowest during the phases of largest
$u_\mathrm{rms}$ and increase as $u_\mathrm{rms}$ decreases. Conversely,
because $\lambda(\vph)$ itself evolves approximately in anti-phase with
$u_\mathrm{rms}(\vph)$, the clustering and Taylor microscale vary in the
same sense: the strongest clustering occurs close to the maximum of
$\lambda$.

The comparison with $\Relam(\vph)$ is qualitatively different.
Although $\langle\sigV\rangle(\vph)/\sigRPP$ varies substantially through
the cycle, $\langle\Relam\rangle(\vph)$ remains nearly constant, with the
small residual modulation identified in
\S\,\ref{sec:singlephase:phase}. Consequently, phases with almost the same
Taylor-scale Reynolds number can exhibit markedly different clustering
intensities. This is a central result of the present measurements:
$\Relam$ does not provide a unique description of the instantaneous
particle-clustering state under periodic forcing.

The strongest correspondence is instead observed with the Kolmogorov time
scale. $\langle\taueta\rangle(\vph)$ varies substantially through the cycle
and approximately in phase with
$\langle\sigV\rangle(\vph)/\sigRPP$: preferential concentration is strongest
when $\taueta$ is largest and weakest when $\taueta$ is smallest. Since
\begin{equation}
  \St(\vph)=\frac{\taup}{\taueta(\vph)},
\end{equation}
the forcing changes the effective particle inertia throughout the cycle even
though the physical properties of the droplets remain fixed.

For the measured Sauter diameter $D_{32}=30.9\,\mu\mathrm{m}$, the particle
response time is $\taup\approx2.91\,\mathrm{ms}$. The Kolmogorov time scale
measured at $\xh=50$ is $\taueta\approx5\,\mathrm{ms}$ at
$\Uo=5\,\mathrm{m\,s^{-1}}$ and $\approx1.4\,\mathrm{ms}$ at
$10\,\mathrm{m\,s^{-1}}$, so that the cycle-averaged Stokes number is
$\St\approx0.56$ and $\approx2.0$, respectively. Over the cycle of T8-5
shown in figure~\ref{fig:phasecluster}, $\taueta$ varies between $3.4$ and
$8.8\,\mathrm{ms}$ and $\St(\vph)$ between $0.32$ and $0.89$; for the
$10\,\mathrm{m\,s^{-1}}$ conditions, $\St(\vph)$ remains between $1.4$ and
$2.7$. At $\Uo=5$ the droplets thus stay on the small-inertia side of the
preferential-concentration maximum throughout the cycle, whereas at $\Uo=10$ they remain on its large-inertia side. In statistically
stationary turbulence, clustering increases with the Stokes number in this
range, up to its maximum at $\St=O(1)$ \citep{Bec2007,Monchaux2012}. For the $\Uo=5$ cases the
observed phase dependence is opposite to this trend, while for $\Uo=10$, where $\St(\vph)>1$, the same trend gives the observed sign: $\sigV$ is largest
when $\taueta$ is largest, that is, when $\St(\vph)$ is smallest. The
phase-resolved clustering at $\Uo=5$ can therefore not be explained by a displacement
of the instantaneous Stokes number along the clustering curve of stationary
turbulence. The same sign was found in turbulence with spontaneously
unsteady large scales, where $\sigV$ reaches its maxima when the
dissipation reaches its minima, for particles with $\St\gtrsim3$
\citep{Zapata2024}.

The measured sequence can therefore be summarised as
\begin{equation}
u_\mathrm{rms}(\vph)\!\downarrow
\quad\Longrightarrow\quad
\eps(\vph)\!\downarrow
\quad\Longrightarrow\quad
\taueta(\vph)\!\uparrow
\quad\mbox{and}\quad
\sigV(\vph)\!\uparrow ,
\label{eq:clustering_chain}
\end{equation}
with the reverse sequence occurring during the opposite half of the cycle.
The links between the carrier-flow quantities follow from the measured
scalings of \S\,\ref{sec:singlephase:phase}, whereas the association with
$\sigV$ is empirical.
This relation should be interpreted as a consistency between the measured
carrier-flow and particle statistics rather than as evidence that a single
quantity alone controls preferential concentration. Nevertheless, it
identifies the evolving small-scale time scale as a much more informative
descriptor of the phase-resolved clustering than $\Relam$.

This distinction is particularly relevant because the standard dissipation
scaling $\eps\sim u_\mathrm{rms}^3/L$ is not recovered over the streamwise
range investigated here. As shown in \S\,\ref{sec:singlephase:phase}, the
substantial phase variations of $u_\mathrm{rms}$, $\lambda$, and $\eps$
largely compensate in their contribution to $\Relam$, leaving
$\Relam(\vph)$ only weakly modulated. The particles, however, remain
sensitive to the resulting variation of $\taueta$. The particle
measurements therefore provide a direct manifestation of the distinction
between a nearly invariant Reynolds number and an evolving instantaneous
small-scale turbulence state.

Finally, the ratio of the particle response time to the forcing period,
$\taup/T=f\taup\leq7.3\times10^{-4}$, is much smaller than unity for all
conditions considered here.
The particle relaxation time is therefore short compared with the forcing
period, and the clustering response is expected to remain close to the
instantaneous state associated with the local value of $\St(\vph)$.
The relatively small phase offsets observed between $\sigV$ and $\taueta$
are consistent with this quasi-static response. At larger $\taup/T$, a finite
response delay would be expected to become increasingly important.

\subsection{Limits of an $\Relam$-based description of clustering}
\label{sec:twophase:interp}

\begin{figure}
  \centerline{\includegraphics[width=\textwidth]%
    {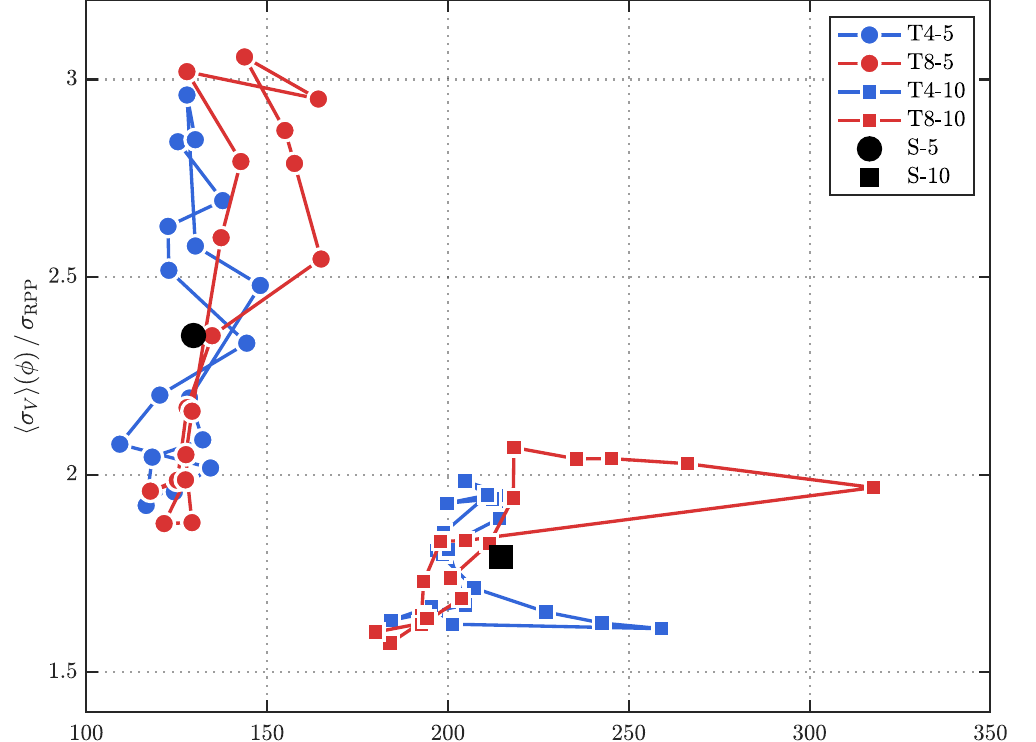}}
  \caption{Phase-resolved trajectories in the
  $(\Relam,\sigV/\sigRPP)$ plane for all four forced conditions. Isolated
  symbols denote the time-averaged steady reference values (S-5 and S-10).
  The trajectories are narrow along $\Relam$ and extended along $\sigV$:
  substantial changes in clustering occur while the Taylor-scale Reynolds
  number varies only weakly.}
  \label{fig:orbit}
\end{figure}

The results of
\S\S\,\ref{sec:twophase:tavg}--\ref{sec:twophase:phase} are brought
together in figure~\ref{fig:orbit}, which shows the phase-resolved
trajectories of the four forced conditions in the
$(\langle\Relam\rangle(\vph),
\langle\sigV\rangle(\vph)/\sigRPP)$ plane. The trajectories are narrow
along the $\Relam$ direction and extended along the clustering direction,
reflecting the weak phase variation of $\Relam$ and the substantially
larger variation of $\sigV$.

The time-averaged steady conditions are included as reference points.
Under periodic forcing, a narrow interval of instantaneous
$\Relam$ corresponds to a broad range of clustering intensities, and the
value of $\sigV$ attained at a given $\Relam$ depends on the forcing phase.
The phase-resolved clustering state therefore cannot be inferred from
$\Relam$ alone.

This result does not imply that $\Relam$ ceases to be a useful turbulence
parameter. Rather, it reflects the fact that $\Relam$ combines
$u_\mathrm{rms}$ and $\lambda$,
\begin{equation}
  \Relam=\frac{u_\mathrm{rms}\lambda}{\nu},
\end{equation}
which vary in opposite directions during the forcing cycle. Their product
therefore remains almost unchanged even though the dissipation and
Kolmogorov time scale vary appreciably. Since $\taueta$ directly enters the
particle Stokes number, two phases with similar $\Relam$ can correspond to
different effective particle inertia and hence different clustering states.

The orbit shown in figure~\ref{fig:orbit} is consequently best understood
as a manifestation of the additional information contained in the
instantaneous small-scale state of the carrier turbulence. In the present
experiments, $\taup/T\ll1$, so the trajectories are traversed in a
quasi-static regime with relatively small phase lag.
In direct numerical simulations of particle-laden turbulence with
spontaneously unsteady large scales, clustering follows the carrier-flow
state with a time lag and describes a hysteresis cycle against $\Relam$
\citep{Zapata2024}. Higher forcing frequencies, larger particles, or other
conditions producing larger $\taup/T$ would provide a natural means
of extending the present measurements towards regimes in which the particle
response no longer follows the carrier-flow variations quasi-statically.

Taken together, the time-averaged and phase-resolved measurements reveal two
distinct effects of periodic forcing. First, forcing systematically increases
the mean preferential-concentration level relative to the corresponding
steady reference state. Second, it produces substantial phase-periodic
variations of clustering that follow the instantaneous small-scale
turbulence quantities despite the weak variation of $\Relam$. The latter
result demonstrates that, in an unsteady particle-laden flow, a complete
description of preferential concentration requires information about the
instantaneous carrier-flow time scales, and not only a Reynolds-number
characterisation.

\section{Conclusions}
\label{sec:conclusions}

We have investigated how slow periodic forcing modifies the turbulence and
preferential concentration of inertial droplets in a spatially developing
turbulent planar jet. The experiments reveal a marked separation between
the response of the mean flow and that of the turbulence and dispersed
phase. Although the forcing frequency is approximately three orders of
magnitude below that associated with the preferred instability of the jet,
and the time-averaged mean-flow development remains essentially unchanged,
the turbulence scales and particle organisation are substantially modified.

For the carrier phase, periodic forcing preserves the self-similar
development of the mean velocity and jet width, but alters the organisation
of the turbulence scales. Most notably, the integral scale is reduced by up
to approximately $40\%$, its downstream growth is strongly inhibited, and
$\Relam$ is reduced by $15$--$20\%$ relative to the corresponding unforced
conditions. The proportionality between the integral scale and jet width is
nevertheless approximately retained, although with a systematically smaller
prefactor under forcing. This result is potentially relevant for turbulence
modelling: while the usual self-similar assumption $L\propto\delta$ remains
valid, the proportionality coefficient is not universal but depends on the
forcing state. Models that prescribe the turbulence length scale solely from
the local jet width may therefore reproduce its downstream scaling while
missing substantial changes in its absolute magnitude under unsteady
conditions. Furthermore, the forced and unforced cases follow the same
time-averaged relationship between the dissipation coefficient and
$\Relam$, with an exponent of approximately $-0.58$. The forcing therefore
changes the turbulence state sampled by the jet without producing a distinct
time-averaged dissipation regime.

The phase-resolved measurements show that this time-averaged description
conceals a pronounced evolution within the forcing cycle. The turbulent
velocity fluctuation and Taylor microscale vary substantially and
approximately in opposite phase, such that $\Relam$ remains within about
$\pm10\%$ of its cycle mean. Similar instantaneous values of $\Relam$ can
therefore correspond to appreciably different values of the dissipation and
Kolmogorov time scale. This demonstrates that $\Relam$ alone does not
uniquely characterise the instantaneous state of periodically varying
turbulence.

The dispersed phase responds strongly to these variations. Periodic forcing
increases the time-averaged preferential concentration and, more
significantly, produces peak-to-peak variations of the Voronoi clustering
intensity amounting to $21$--$49\%$ of its cycle mean. Clustering is
systematically enhanced during the part of the cycle characterised by weaker
dissipation and a larger Kolmogorov time scale. Because the particle response
time is fixed, this variation changes the instantaneous particle Stokes
number.
For the present droplets, with a measured Sauter diameter of
$30.9\,\mu\mathrm{m}$, $\St$ varies through the cycle
($\St\approx0.32$--$0.89$ at $\Uo=5\,\mathrm{m\,s^{-1}}$ and
$1.4$--$2.7$ at $10\,\mathrm{m\,s^{-1}}$), so that clustering is strongest
when the droplets are least inertial. For the $\Uo=5$ droplets this is opposite to the
Stokes-number dependence of clustering in statistically stationary
turbulence, and shows that the instantaneous Kolmogorov-scale Stokes number
alone does not account for the clustering response. The phase-resolved trajectories
therefore demonstrate that markedly different particle distributions can
occur at similar values of $\Relam$: the particle field provides a sensitive
response to changes in the small-scale turbulence that are largely hidden
when the carrier flow is described by a single Reynolds number.

These observations suggest that slow temporal modulation may provide a means
of manipulating particle organisation without comparably altering the mean
transporting flow. In the present regime, the particle response time is much
shorter than the forcing period, and the clustering consequently follows the
evolution of the carrier turbulence with relatively small phase offsets.
Forcing at higher frequencies, or particles with longer response times,
should introduce stronger memory and phase-lag effects; a time lag between
clustering and the carrier-flow state has been reported in turbulence with
spontaneously unsteady large scales \citep{Zapata2024}.
Exploring this regime would establish whether both the
magnitude and phase of preferential concentration can be controlled through
the forcing parameters.

The present particle measurements are restricted to $\xh=50$, and the
accessible forcing amplitudes and frequencies are constrained by the
facility. In addition, the dissipation is inferred from hot-wire measurements
rather than from direct spatial measurements of the velocity gradients, and
the droplet size distribution was measured at the outlet of an injection
nozzle rather than at the measurement station.
Within these limitations, the experiments establish that modest periodic
forcing can substantially reorganise turbulence scales and produce pronounced
changes in inertial-particle clustering while leaving the time-averaged
mean-flow development almost unaffected. This separation between mean-flow
transport and particle organisation provides a physical basis for using
temporal modulation as a control parameter in particle-laden turbulent
flows.       

    

    

\begin{bmhead}[Declaration of interests.]
  The authors report no conflict of interest.
\end{bmhead}

\begin{bmhead}[Data availability statement.]
  The data supporting the findings of this study are available
  from the corresponding author upon reasonable request.
\end{bmhead}

\begin{bmhead}[Author ORCIDs.]
  M.~Ahmad, \url{https://orcid.org/0000-0001-9823-6051};
  C.~Cuvier, \url{https://orcid.org/0000-0001-6108-6942};
  M.~Obligado, \url{https://orcid.org/0000-0003-3834-3941}.
\end{bmhead}

    

\bibliographystyle{jfm}
\bibliography{references/references}

\end{document}